# Comparison of mechanical properties of $Ag/W_{1-x}Ti_xB_{2.5}$ and pure silver coatings deposited by PLD/HIPIMS method

Katarzyna Zielińska[1*], Mateusz Włoczewski[1,2], Rafał Psiuk[1], Jacek Hoffman[1], Ewa Wojtiuk[1], Piotr Bazarnik[2], Tomasz Mościcki[1]

[1] Institute of Fundamental Technological Research POLISH ACADEMY OF SCIENCES ul. Pawińskiego 5B 02-106 Warsaw, Poland
[2] Warsaw University of Technology, Faculty of Materials Science and Engineering, Warsaw, Poland

*Corresponding author: kzielin@ippt.pan.pl, tmosc@ippt.pan.pl

## Abstract:

Transition metal borides are attracting increasing interest due to their unique properties. They are not only characterised by very high hardness, but also considerable chemical and thermal stability. Silver, on the other hand, is a good material for increasing electrical and thermal conductivity, wear resistance and has antibacterial properties due to its biological characteristics. Combining these two materials can provide superhard bilayers with increased functional properties. In this study, it was decided to synthesise $Ag/WB_{2.5}$, $Ag/W_{0.76}Ti_{0.24}B_{2.5}$ coatings and compare their properties to the individual components. The silver coating was produced by pulsed laser deposition (PLD), while the $WB_{2.5}$ and $W_{0.76}Ti_{0.24}B_{2.5}$ coatings were formed by high-power pulsed magnetron sputtering (HiPIMS). To determine the mechanical properties, nanoindentation tests, adhesion of the coatings by scratch -test and wear resistance by abrasion in reciprocating motion were tested. In all cases, the silver film contributed to an increase in the wear resistance of the materials without major changes in the hardness results of the materials. In addition, the $Ag/W_{0.76}Ti_{0.24}B_{2.5}$ film showed very good adhesion to the substrate. Human hand wiping simulator was also carried out using - Tribotouch. After 36 000 cycles $Ag/W_{0.76}Ti_{0.24}B_{2.5}$ coating was slightly deformed, which was not visible macroscopically. This result is more than three times greater than for the pure silver film. It was also decided to carry out corrosion tests in an environment of 0.9% NaCl. The $Ag/W_{0.76}Ti_{0.24}B_{2.5}$ bilayer has very good corrosion resistance, similar to pure silver.

# 1. Introduction:

In recent years, there has been a high demand for functional coatings that are primarily characterised by high wear resistance and mechanical stability. These types of hard, functional coatings represent a new generation of coatings [1]. Unfortunately, the main problem associated with the use of functional layers is their limited hardness and, consequently, low tribological wear resistance and insufficient stability at elevated temperatures. It is worth noting that wear is one of the main factors causing high energy consumption and significant economic losses. Reducing the wear of interacting surfaces is essential to extend their service life [2].

One of the materials gaining increasing popularity are transition metal borides (TMBs). Compared to films made of nitrides or carbides, TMBs have even higher hardness due to the higher electron density of transition metals and strong covalent bonds of light elements [3]. In addition, they are characterised by a high melting point, chemical and thermal stability [4], as well as high thermal and electrical conductivity [5]. These unique properties make it possible to use these materials in very demanding and diverse environments, from wear-resistant tools [6] to nuclear fusion devices [7], as well as in biomedical applications [8]. Transition metal borides coatings are usually produced by physical vapour deposition (PVD). Magnetron sputtering is a particularly common technique for depositing these materials. It is one of the most versatile methods for depositing high-quality layers and can be used for industrial coating of various substrates. It uses a power source with both direct current (DC) [4, 9], RF [10, 11] as well as high power impulse magnetron sputtering HiPIMS [11]. Of these three, HiPIMS uses short (tens to hundreds of µs) but highly energetic power pulses, generating a dense plasma with a high degree of ionization. A key advantage of this technique is the increased ionization fraction of the sputtered species, which enables ion-assisted film growth, in contrast to conventional magnetron sputtering. This leads to enhanced adatom mobility and increased ion bombardment at the growing film, resulting in coatings with improved density, higher hardness, and better adhesion to the substrate [12, 13]. Unfortunately, the main limitation of this type of material is its relatively high Young's modulus, which directly translates into increased brittleness and susceptibility to cracking [11, 14]. This problem can be partially overcome by appropriately alloying tungsten borides with other transition elements, such as tantalum [11, 15], aluminum [16] or titanium [17, 18]. The addition of these metals allows for the modification of the coating structure, which in turn leads to an improvement in their mechanical properties and increased resistance to brittle fracture [18]. However, it is worth noting that despite many results achieved in improving the mechanical properties of transition metal borides, they still belong to a relatively new class of materials. Therefore, there are a limited number of publications on their functionalisation.

Silver, on the other hand, is a very good material in terms of protection against bacteria and improving sliding properties [19, 20, 21]. There are many silver-based materials, including silver nitrate and silver sulfadiazine, which are commonly used in medical treatment [22]. Due to these distinctive biological properties, the production of composite layers containing silver is attracting increasing interest [23, 24, 25]. In one study, M. Vidiš et al. decided to produce a single-layer $Ag/TiB_2$ nanocomposite coating deposited by conventional magnetron sputtering [21]. Small clusters of Ag were deposited in extremely hard $TiB_x$, which acted as a protective matrix. Even less than 0.1% at. of silver content contributed to a significant inhibition of bacterial growth. As the amount of Ag increased, this effect became stronger, reaching 97% inhibition of E. coli bacterial growth for a sample containing 24% at. Ag. Despite a significant decrease in hardness with increasing silver content, the coating exhibits good wear resistance. Additionally, silver acts as a solid lubricant, reducing the friction coefficient from 0.77 to 0.35. In the work by B. Chauhan et al. [26], it was decided to deposit Ni–Ag coatings with various Ag content on the Cu surface using an electrodeposition technique. Increasing the Ag content reduces the coefficient of friction (CoF) compared to the initial run-in period, and this effect is maintained throughout subsequent sliding. The wear rate remained low for all Ag concentrations relative to the copper substrate, due to the effect of Ag acting as a lubricant material within the coatings. W. Aperador and others [27] investigate the development of silver (Ag)-doped zirconia ($ZrO_2$) coatings deposited on stainless steel using magnetron sputtering technique. The results demonstrate that Ag doping, combined with increased oxygen content, significantly improves both wear and corrosion resistance. The optimized Ag-$ZrO_2$ coatings developed in this study demonstrate potential for many applications, where enhanced durability and corrosion resistance are critical. In another study, Ag-DLC films prepared by dual PLD were produced [28]. Despite a decrease in hardness with increasing silver content, studies have shown that the layers produced are potentially useful biomaterials with antimicrobial properties.

In turn, Zhang et al. devised an innovative method combining mechanical alloying (MA) and laser powder bed fusion (LPBF) of silver and nickel powders in austenitic stainless steel. Thanks to dispersed Ag-rich nanoparticles, the samples exhibited antibacterial activity, killing 99.999% of Escherichia coli and Staphylococcus aureus bacteria within 6 hours [29]. Fiantok et al. decided to produce $ZrB_2$ thin films alloyed with silver prepared by magnetron co-sputtering [30]. An improvement in tribological properties was observed, with the coefficient of friction decreasing from COF ($ZrB_{2.31}$) ~ 0.9 to a value of COF ($Zr_{0.26}Ag_{0.74}B_{0.89}$) ~ 0.25. With increased Ag content, values of Young's modulus decrease from $E_{ZrB2.31}$ = 375 GPa to E ($Zr_{0.26}Ag_{0.74}B_{0.89}$) = 154 GPa, unfortunately this was associated with a decrease in hardness from H ($ZrB_{2.31}$) = 30 GPa to a value of H ($Zr_{0.26}Ag_{0.74}B_{0.89}$) = 4 GPa. It is also worth noting that silver-based coatings are well known for their high thermal and electrical conductivity. In durable composite coatings, silver is often used as a conductive phase introduced into a ceramic matrix [31] or into advanced materials based on carbon nanotubes [32]. This makes it possible to maintain high mechanical strength while ensuring effective heat dissipation and electrical conductivity.

Athough, there are currently many studies on both experimental and theoretical research on tungsten borides, but there are no studies yet on their functionalisation through silver modification. The main objective of this work is to create functional Ag/$W_{1-x}Ti_xB_{2.5}$ bilayers by combining two methods, i.e. MS - PLD, and to compare their properties with a layer composed of pure silver. The design of a coating consisting of a soft film of silver with lubricating properties, combined with the high mechanical strength of borides, is primarily intended to increase wear resistance. The deposition methods used ensure the production of highly ionised, high-energy plasma, which in turn leads to denser layers with fewer defects and stresses. At the same time, silver particles generated during ablation (increased roughness) will be used in an attempt to distribute them in the tungsten boride matrix. The formation of silver droplets of various sizes, sometimes larger than the thickness of the coating, will reinforce the material. Furthermore, during abrasive wear, the areas with the highest roughness may break off as a result of friction, and the exposed silver will act as a lubricant. In turn, the combination of the properties of tungsten borides, such as high hardness, wear and abrasion resistance, and the properties of silver, is expected to enable the use of bilayers in various fields, including the medical automotive, space and aviation industries. Nanocoatings could be used in the production of surgical instruments such as scissors, scalpels and needles, but also in places of mass use, e.g. on door handles and lift buttons. The silver layer could have a positive effect on the lubricating properties of the nanocoatings and can additionaly improve their biological performance by providing good corrosion resistance.

# 2. Experimental details

## 2.1 Materials and process parameters

In order to develop the desired coatings, three targets were used:
- commercial one-inch target Ag (Kurt J. Lesker Company),
- two-inch target $WB_{2.5}$: tungsten (purity: 99.9%, particle size: 25 µm), amorphous boron (purity: 95%, particle size: 1 µm),
- two-inch target $W_{0.76}Ti_{0.24}B_{2.5}$: tungsten (purity: 99.9%, particle size: 25 µm), titanium (purity: 99.8%, particle size: 250 - 350 µm), amorphous boron (purity: 95%, particle size: 1 µm).
The targets composed of $WB_{2.5}$ and $W_{0.76}Ti_{0.24}B_{2.5}$ were synthesised by SPS (spark plasma sintering) in accordance with the procedure described in the work of T. Mościcki et al [5]. In the first stage of the study, Ag films were deposited using the PLD (Pulsed Laser Deposition) method. As the substrate were used monocrystalline silicon samples with dimensions of 10 x 10 x 0.7 mm. The samples were placed in a vacuum chamber. The process was carried out using an Nd: YAG laser (Quantel, Fraction). The laser beam was focused on a rotating target, positioned at an angle of 45° to the incident beam. The target was located parallel to the sample substrate at a distance of 40 mm. Detailed process parameters are summarised in .

Table 1. The use of relatively high fluence was intended to produce a rough surface covered with silver particles, which, after the magnetron layer was deposited, would act as a "reinforcement" for the boride matrix. The expected effect was to increase abrasion and corrosion resistance, as well as improve the thermal and electrical conductivity of the resulting bilayer.

*Table 1 Silver deposition process parameters.*

| Power [W] | Laser fluence [J/cm$^2$] | Wavelength [nm] | Pulse duration [ns] | Repetition ratio [Hz] | Deposition time [min] | Pressure [Pa] |
|---|---|---|---|---|---|---|
| 10 | 2.52 | 1064 | 10 | 10 | 30 | 5 x 10$^{-5}$ |

The next step was to produce further coatings on top of the previously produced silver films using the HIPIMS (High Power Impulse Magnetron Sputtering) method. This process was carried out using the magnetron gun PREVAC MS2 63C1, which allows precise control of the deposition parameters. Two separate deposition processes were performed using identical parameters, but with different targets: $WB_{2.5}$ and $W_{0.76}Ti_{0.24}B_{2.5}$. The process was carried out in a vacuum chamber, which was evacuated to a base pressure of $2 \times 10^{-6}$ Pa and then filled with argon to a working pressure of 0.9 Pa. The distance between the magnetron target and the substrate was 8 cm. Prior to the actual process, a pre-sputtering of the targets was carried out for 5 min with the substrate covered. The purpose of this process was to clean the surface of the magnetron targets from possible contaminants. The detailed parameters of the deposition process used are summarised in Table 2. The parameters were selected based on the study by Mościcki et al. [18], in which hard and wear-resistant $W_{1-x}Ti_xB_{2.5}$ coatings were obtained at a deposition temperature of 400 °C.

*Table 2 Magnetron deposition process parameters.*

| Shield composition | Power [W] | Repetition ratio [Hz] | Pulse duration [µm] | Temperature [°C] | Deposition time [h] | Gases | Pressure [Pa] |
|---|---|---|---|---|---|---|---|

| $WB_{2.5}$ | 250 | 700 | 200 | 400 | 1 | Argon | 0.9 |
|---|---|---|---|---|---|---|---|
| $W_{0.76}Ti_{0.24}B_{2.5}$ | | | | | | | |

### 2.2 Characterisation of morphology and microstructure

A Joel scanning electron microscope was used to analyse the surface morphology. This instrument is additionally equipped with Energy Dispersive X-ray Spectroscopy (EDS), which made it possible to analyse the chemical composition of the surfaces of the materials studied. The analysis was carried out at an accelerating voltage of 10 kV. However, due to the limited accuracy of standard SEM-EDS methods for analysing the chemical composition of light elements such as boron, TOF-ERDA method was used to determine the elemental composition. The analysis was performed using $^{127}I^{7+}$ ions with an energy of 23 MeV, incident on the sample surface at an angle of 20°, while the TOF-ERDA detector was angled at 37.5° to the beam axis. In order to obtain more accurate characterisation of the samples, a FESEM-FIB ZEISS Crossbeam 350 was used. Using the Focused Ion Beam, a cross-sectional view of the samples was obtained, which enabled direct measurement of the thickness of the individual coatings. Subsequently, a decomposition of the elemental map was performed, which allowed the determination of the chemical composition in cross-section and the distribution of the individual components in the material structure. For a more detailed analysis of the microstructure, it was also decided to perform TEM studies. The samples were applied to a standard copper TEM grid coated with a layer of amorphous carbon. After the grid had completely dried, the samples were placed in the microscope. The tests were carried out on an FEI Titan Cubed 80–300 TEM microscope with an accelerating voltage of 300 kV. General images were recorded in bright field TEM mode. Phase composition analysis was carried out using a Bruker D8. The study used $CuK\alpha_1$ radiation with a wavelength of $\lambda$ = 1.5406 Å. Due to the need to analyse very thin films, a measurement geometry for an incidence angle of the X-ray beam of 2° was used. The diffraction spectra were recorded at 40 kV and a current of 40 mA in the 2Θ angle range from 20° to 75°, with a measurement step of Δ2Θ = 0.025° and a count time of 5s. The analysis of the surface chemical composition of the samples was performed using the XPS/AES Microlab 350 spectrometer from Thermo Electron. A non-monochromatic X-ray source (Al $K_\alpha$) with an energy of 1486.6 eV and a power of 300 W was used for XPS measurements. Survey spectra were recorded in a wide energy range from 1350 eV to 0 eV with a step of 1.0 eV at the transition energy CAE = 100 eV, while high-resolution spectra (HR-XPS) were collected in a narrow range of binding energies with a step of 0.1 eV, with a transition energy of CAE = 40 eV. All XPS spectra for individual samples were measured with a horizontal resolution of 0.2 $cm^2$. The recorded HR-XPS spectra were deconvoluted using Thermo Avantage software (version 5.9911) from Thermo Fisher Scientific, using the Gauss-Lorentz function at a constant ratio G/L = 0.35 (± 0.05). The background was corrected using the Smart model. The correction of the results due to the "surface charging" effect was carried out in relation to the carbon peak C 1s = 285.0eV. The surface roughness was measured using a Sensofar Tech Industries interferometric profilometer. Four independent measurements were taken for each sample, from which the arithmetic mean deviation of the profile from the mean line (Ra) was calculated.

### 2.3 Mechanical and tribological properties

Nano-indentation of prepared coatings was measured with the use of in-situ SEM Alemnis nanoindenter equipped with the diamond Berkovich tip. In order to obtain higher measurement accuracy, before measurements, CSM (Continous Stiffness Measurement) calibration at reference fused-silica sample was performed. Three CSM indentations were performed at the center of each sample, with a minimum spacing of 100 um between indents. The CSM technique has been described in detail in [33, 34, 35]. Briefly, each test was initiated approximately 250 nm above the sample surface and penetrated up to 50 nm without oscillation to get a clean initial contact. Next, a sinusoidal oscillation was superimposed on the nominally increasing load, with a frequency of 2 Hz, an amplitude of 20 nm, and a loading rate of 10 nm/s, until a maximum load of 25 mN was reached. At the end, the sample was unloaded to 0 nm without oscillation. In order to eliminate the influence of temperature drift and apparatus drift, the measurement data was corrected accordingly using dedicated software. Analysis of the results was carried out using specialised software provided by the nanoindenter manufacturer. All values are reported as mean ± standard error based on independent measurements. The assessment of the adhesion of coatings to the substrate is also significant in subsequent applications. A scratch test was carried out using the Micro Combi Tester MCT3. The test involves scratching the surface of the sample with an indenter of known geometry. A Rockwell diamond indenter with a rounding radius of 0.2 mm was used as a counter-sample. Moving the specimen at a constant speed in the direction perpendicular to the blade, the specimen was scratched by applying a set force, causing a recess. The length of the scratch was 3 mm, the load increased linearly between 0.03 N and 10 N, and the feed rate was 6 mm/min. During the test, corresponding graphs were created by continuously recording the pressure force, blade penetration depth, friction force and acoustic emission signal associated with coating cracking, among others. These data were correlated with microscopic analysis of the scratches made, carried out using a Nikon Eclipse LV150N optical microscope. On the basis of the results, the adhesion of the coatings to the substrate was assessed and the critical forces at which layer detachment occurred were determined.

Wear resistance was assessed by means of an abrasion test in reciprocating motion, carried out using the Micro Combi Tester MCT3. The test used an $Al_2O_3$ alumina ball with a diameter of ϕ = 6 mm, which moved across the surface of the specimen in a cyclic manner. A constant load of 0.5 N was applied to all test specimens. The movement of the ball was at a linear speed of 1 cm/s, and the total sliding distance was 10 m. Subsequently, the material loss per unit area was determined using an optical profilometer (Sensofar Tech Industries) based on four roughness profiles measured at the abrasion site for each sample. The results are reported as mean ± standard error calculated from n independent measurements. In addition, it was also decided to analyse the chemical composition of the abrasion marks. The elemental map was performed in order to identify any changes in the material.

The fracture toughness of the coatings was assessed by performing an indentation fracture toughness test, which involves penetrating the surface of the sample with an indenter. This process causes a local increase in stress, which leads to deformation and crack initiation in the coating. The test was carried out using The Micro Combi Tester MCT3 hardness tester, equipped with a Cube Corner diamond indenter. This indenter is characterised by a 35.3° angle between the symmetry axis and the wall surface, resulting in more than three times the volume of material displaced compared to a classic Berkovich indenter. As a result, significantly higher stresses and strains are generated, allowing them to generate cracks at lower loads [36]. Due to different deformation mechanisms, this method is applied to ceramic materials. Two maximum loads of 200 mN and 300 mN were applied during testing. Four impressions were made on each specimen for each load. The materials were then subjected to microscopic observations using a Hitachi Su-70 scanning electron microscope to accurately analyse the appearance, distribution and amount of damage created on the imprints. The resistance to brittle fracture $K_C$ was determined based on the following formula:

$$K_C = \delta \left(\frac{E}{H}\right)^{1/2} \frac{F_m}{c^{3/2}}$$

where $F_m$ is the maximum load (200 mN and 300 mN). In the equation: E is the modulus of elasticity, H is the hardness, c is the crack length, and δ is an empirical constant depending on the shape of the indenter. For the cube-corner indenter used, a value of δ=0.036 was assumed.

To assess hand abrasion resistance, finger abrasion was simulated using the Tribotouch device. This device allows an accurate representation of the chemo-mechanical system of the human hand, according to international standards. The touch of the human hand consists of two phases: a mechanical impact and a subsequent pushing and sliding force, as illustrated in Figure 1. Tests were carried out on specimens containing Ag and Ag/$W_{0.76}Ti_{0.24}B_{2.5}$, using standard parameters in accordance with EN 60068-2-70. The experiment used a test material with a properly roughened and structured surface to simulate the properties of human, clean skin, according to the specifications of the ISO 12947-1/IEC 68-2-70 s. The force at which the sample was struck was 5 N, the frequency 2 Hz, the friction path 4 mm and the angle of the simulated finger relative to the sample was 45°. Surface appearance is crucial to the perceived quality of the material. Therefore, the number of cycles after which the first visible changes started to appear on the surface of the samples was analysed. After testing, the samples were subjected to detailed microscopic observations using a Hitachi Su-70 scanning electron microscope to determine any surface degradation. The chemical composition at the imprint sites was also investigated. These results provided a qualitative assessment of the abrasion resistance of the coatings.

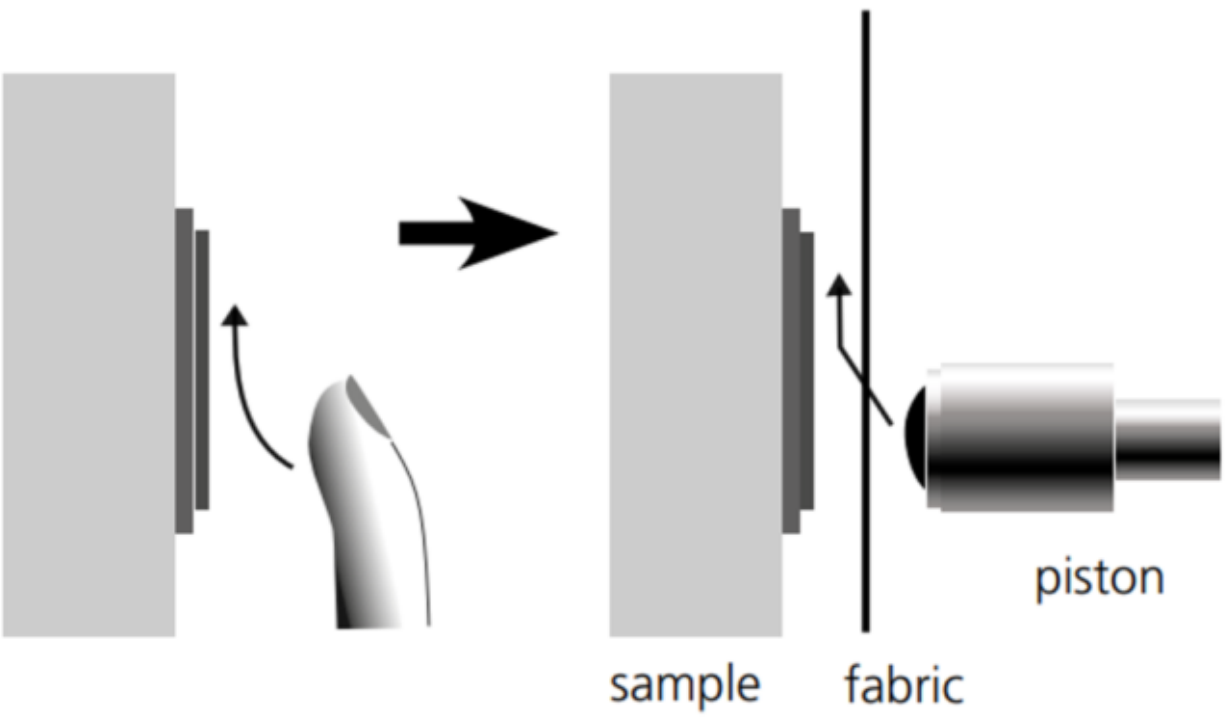

*Figure 1 Principle of human touch [37].*

## 2.4 Corrosion properties

The corrosion resistance of the films produced was also investigated. The role of the electrolyte was a solution of 0.9% NaCl, simulating the approximate working environment of the tested materials. A three-electrode system was used, in which the working electrode was the test sample, the reference electrode was a saturated calomel electrode, while a platinum wire served as the auxiliary electrode. First, current-free open-circuit potential measurements were performed for 120 minutes. This is a non-destructive test that is necessary for further research. By analyzing the change in potential over time, a value close to the corrosion potential was determined for each sample. Then one of the most common methods in electrochemical testing, the potentiodynamic method, also known as anodic polarization curves, was carried out. The results are obtained by polarizing the sample with time-varying potential. Polarization took place in the range: - 250 mV to 250 mV from $E_{ocp}$ (open circuit potential), step 0.5 mV/s. In order to obtain adequate statistics and correct results, three measurements were made on each sample. From the obtained polarization curves, the characteristic average electrochemical values were determined: $i_{kor}$ - corrosion current density, $E_{kor}$ - corrosion potential. Using the linear polarization method, the polarization resistance $R_{pol}$ was also determined.

# 3. Results and discussion:

## 3.1 Morphology and microstructure

### 3.1.1 Surface and cross – section

Figure 2 shows images of surface topography obtained by scanning electron microscopy (SEM) for the three test samples. After the silver PLD deposition process, the surface shows relatively irregularly distributed objects marked with arrows (Figure 2a)). Most likely, these are silver particles that were deposited on the surface as a result of the laser ablation process. A strong, short-duration laser pulse locally heats the surface of the target, causing larger fragments of molten or semi-molten material to be ejected. These particles are spherical in shape and vary in size from less than one micrometer to as much as 5 µm. After the deposition process of $WB_{2.5}$ and $W_{0.76}Ti_{0.24}B_{2.5}$ coating on the previously applied silver film, no significant changes were observed on the surface. Spherical-shaped particles are still visible (Figure 2 b), c)). However, some of the smaller particles disappeared or became less distinct. Most likely, these particles were small enough to be completely covered by the newly applied $WB_{2.5}$ and $W_{0.76}Ti_{0.24}B_{2.5}$ coatings. In order to analyse the structure in more detail, Figure 3 shows high resolution TEM images of coatings. The image of silver film shows parallel contrast lines with regular periodicity, which are the result of phase and diffraction effects (Figure 3 a)). The interference fringes are related to the atomic arrangement of the silver crystal. The presence of these bands confirms the high crystallinity of the analysed coating. $W_{0.76}Ti_{0.24}B_{2.5}$ In the case of the $W_{0.76}Ti_{0.24}B_{2.5}$ coating, no grains or columnar structures were observed (Figure 3 b)). The material has an amorphous structure, which is confirmed by the results of XRD analysis (Figure 7).

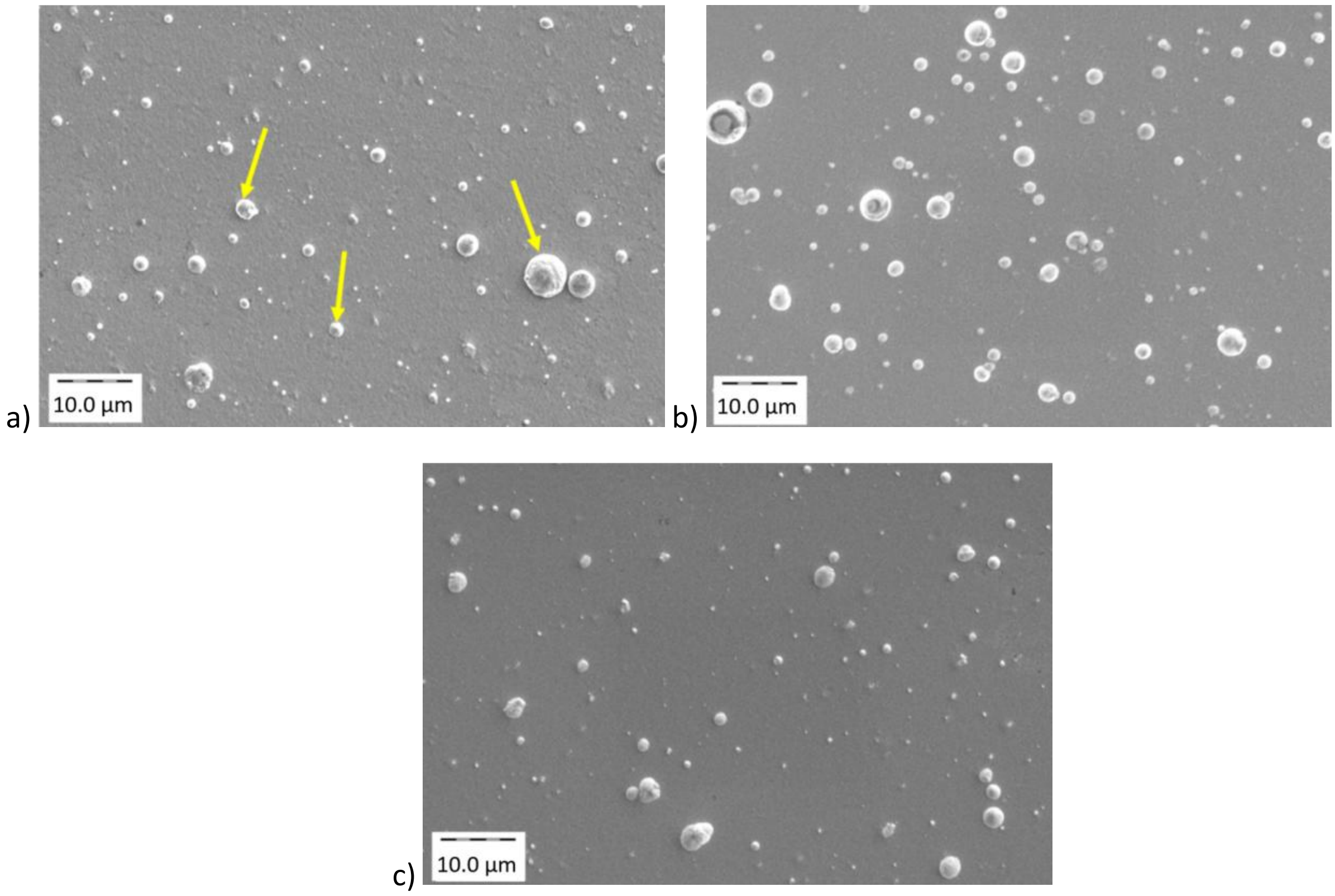

*Figure 2 SEM images of the surface of a) Ag, b) Ag/$WB_{2.5}$, c) Ag/$W_{0.76}Ti_{0.24}B_{2.5}$; yellow arrows indicate silver particles produced by laser ablation.*

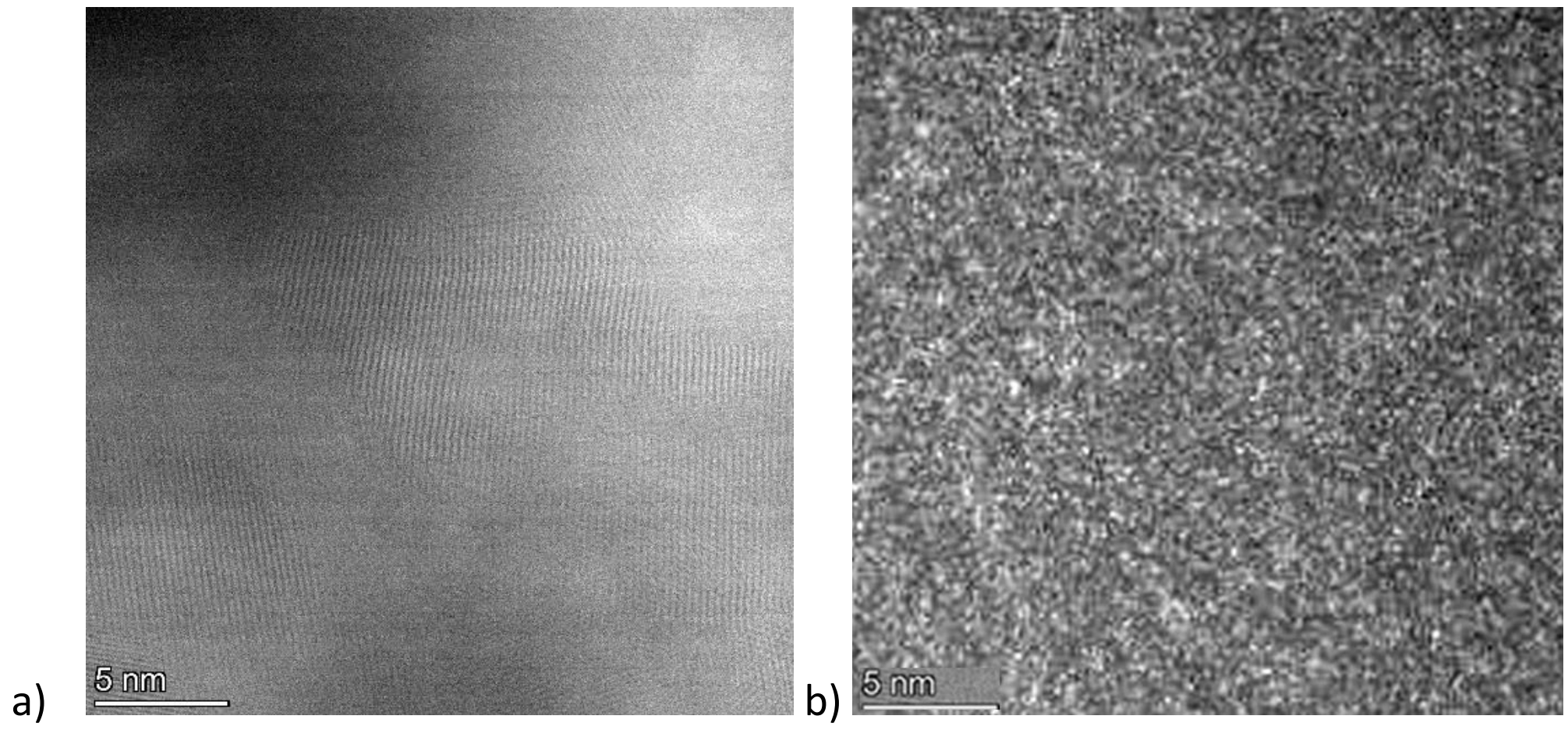

*Figure 3 Structural analyses of a) Ag and b) $W_{0.76}Ti_{0.24}B_{2.5}$ coating, based on HR-TEM images.*

To measure the thickness of the coatings, the samples were etched using a FIB beam. The resulting images of cross sections from a scanning electron microscope are shown in Figure 4. The surface of the PLD film is highly developed, and a silver particle can be seen in cross-section in the image (Figure 4 a)). The average thickness of the silver layer is 530 nm. The high roughness is due to the fact that the phase explosion threshold is exceeded, so that the silver particles are also deposited on the substrate of the material. After the $WB_{2.5}$ coating was applied to the silver by the MS method, there was a significant change in the appearance of the material (Figure 4 b)). The surface is noticeably smoother than for pure silver. A uniform coating of borides completely covers the silver substrate. However, larger silver particles protruding above the surface are still visible. The thickness of the film has increased by more than 1.5 µm, reaching an average value of 2.4 µm. Unfortunately, already at the initial stage of testing, a microscopic crack can be observed at the interface between the layers. This is due, among other things, to the large difference in mechanical properties (Young's modulus and hardness) and different thermal expansion. Silver is soft and susceptible to deformation, unlike the boride coating. This combination promotes the formation of cracks along the interface. In the case of the WTiB film, a similar structure is observed, but without noticeable discontinuities in the structure (Figure 4 c)). The addition of titanium reduced the Young's modulus, decreasing the brittleness of the material [18]. The coating is uniformly deposited, and larger silver particles can still be seen on the surface. The average thickness of the bilayer coating is 2.35 µm.

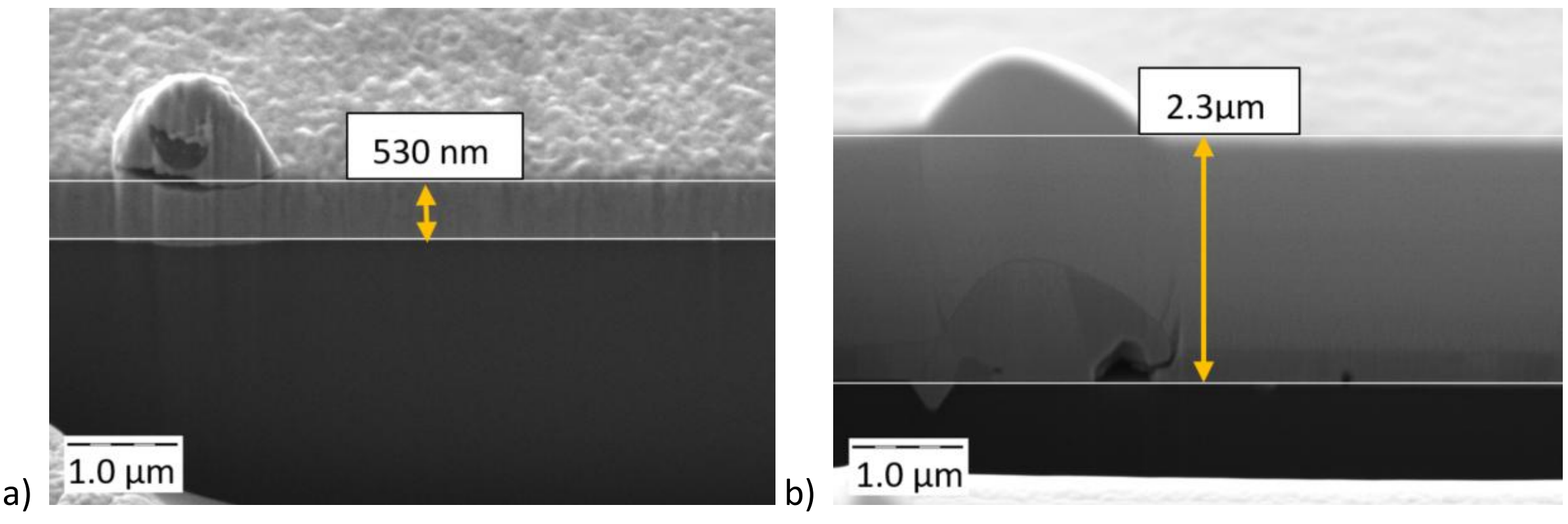

c) 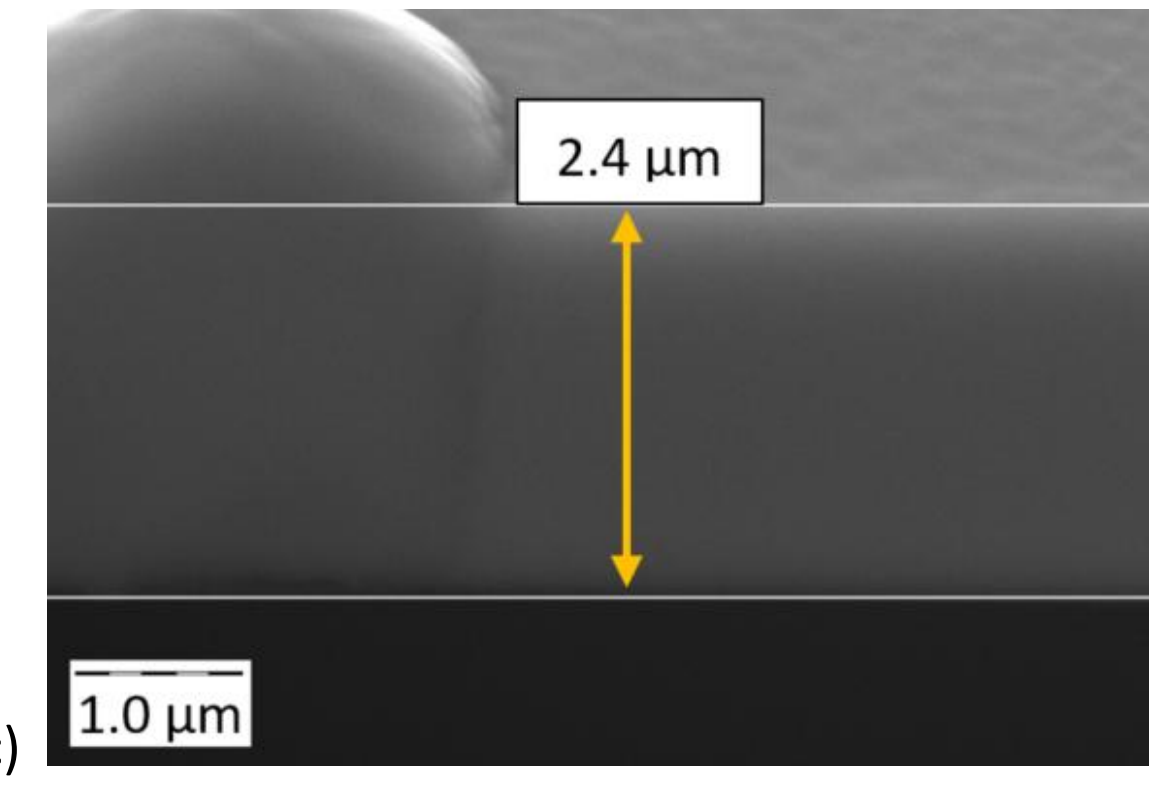

*Figure 4 SEM images of the cross - section of a) Ag, b) Ag/$WB_{2.5}$, c) Ag/$W_{0.76}Ti_{0.24}B_{2.5}$.; magnification 17 000x.*

It was also decided to check the chemical composition on a cross - section of a sample composed of Ag/$W_{0.76}Ti_{0.24}B_{2.5}$., also including a cross-section of one of the particles. Figure 5 shows images of the elemental map distribution. The substrate of the sample, as expected, consisted solely of silicon. The deposited silver is visible over the entire surface, but the film is very thin. The largest accumulation of silver is in the particle, visible in the initial photo. It is due to fact that particle was produced by the PLD method and consists of pure silver. The coating produced by magnetron sputtering covers the entire surface, including the silver particle. The elements W, Ti, B are evenly distributed over the surface. The picture shows that the intensity of the elements tungsten and titanium is higher relative to the boron content, but it is still detectable.

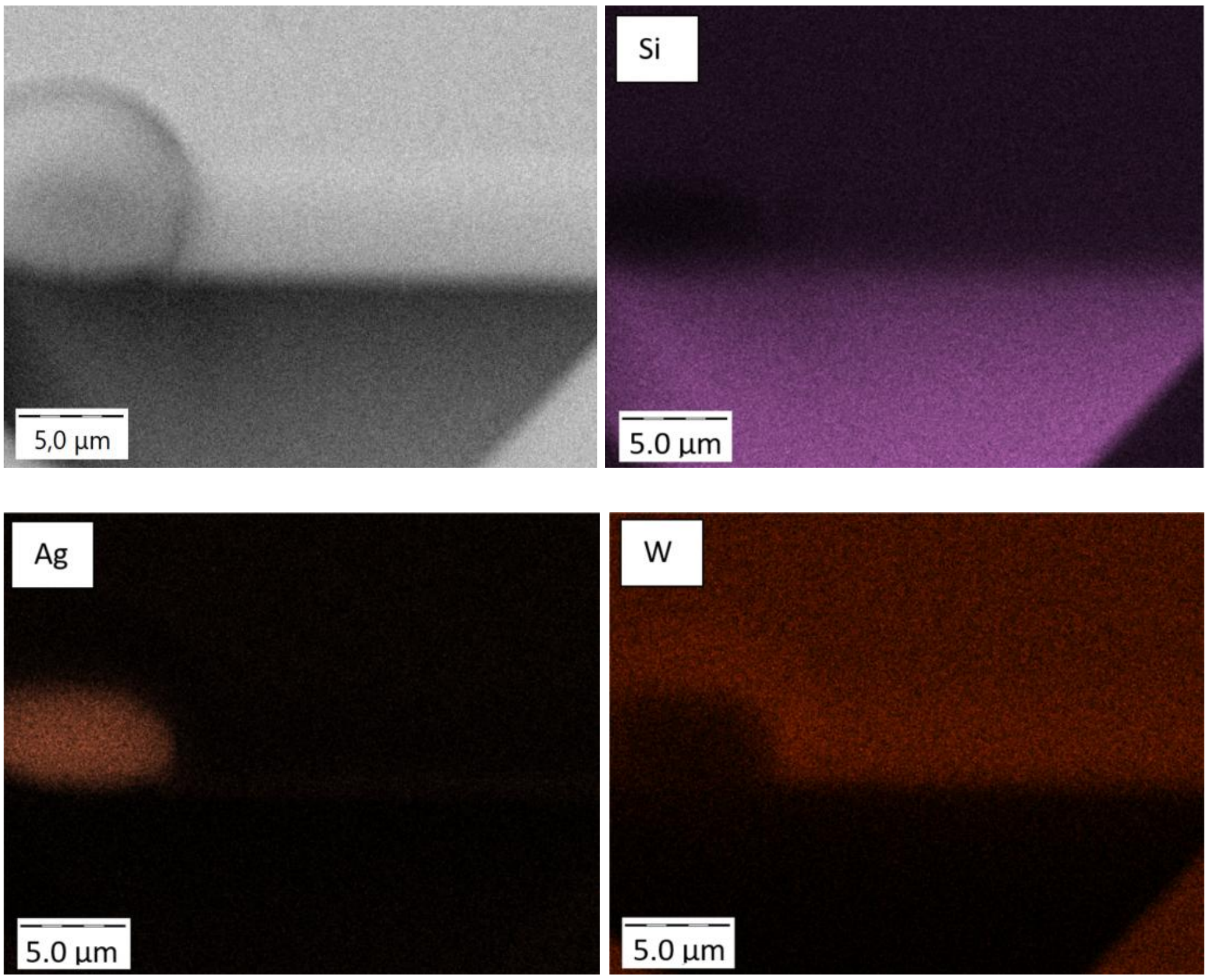

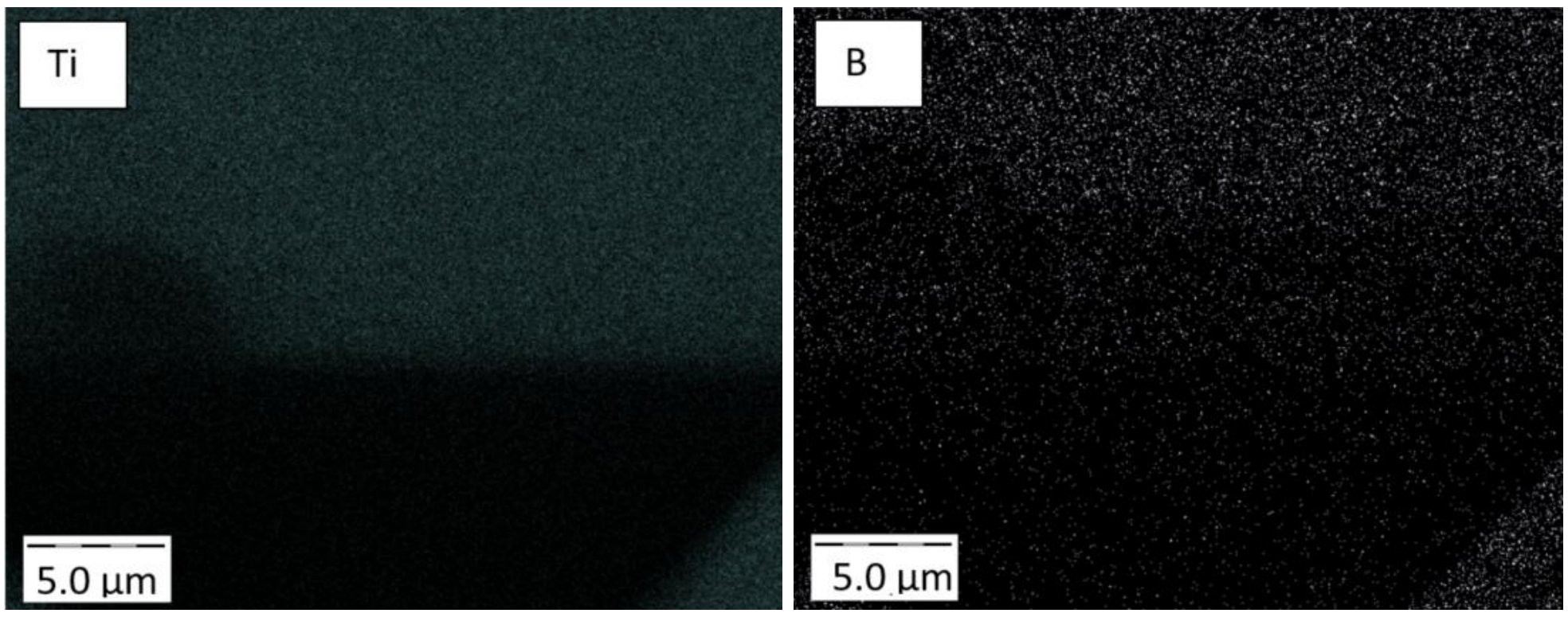

*Figure 5 Distribution of the map of elements of coating Ag/$W_{0.76}Ti_{0.24}B_{2.5}$.*

The chemical composition of the materials on the surface was also investigated. The high atomic number and intense characteristic lines in the EDS spectrum for silver allow for its unambiguous identification. Thus, the PLD-only film was examined using the standard SEM - EDS method. As expected, regardless of the measurement location, it consists exclusively of silver. Unfortunately, the EDS chemical composition test method has some limitations regarding the measurement of light elements such as boron. Measurements are particularly difficult in the presence of high atomic weight elements (in this case tungsten). Boron, due to its very low characteristic radiation energy (Kα boron about 0.183 keV), has low signal intensity and is susceptible to absorption by the EDS detector and scattering by heavy elements [38, 39]. For this reason, the Ag/$WB_{2.5}$ and Ag/$W_{0.76}Ti_{0.24}B_{2.5}$ bilayer were examined using the ToF-ERDA method. Figure 6 a), b) shows the average values obtained for the content of each element with the measurement error. As assumed, in both cases, the layers consist primarily of boron. For $WB_{2.5}$ layer the theoretical atomic composition is approximately 71.4% at. B and 28.6% at. W. ToF-ERDA results showed approximately 65% at. B and 31% at. W, which is in good agreement with the expected stoichiometry. For the $W_{0.76}Ti_{0.24}B_{2.5}$ coating, the theoretical composition corresponds to approximately 71.4% at. B, 21.7% at. W and 6.9% at. Ti, while ToF-ERDA measurements yielded about 68 ± 5% at. B, 22 ± 1% at. W and 6.8 ± 0.7% at. Ti. These results are very close to the expected values and confirms that the deposited coating composition is consistent with the intended stoichiometry.

In both the Ag/WB and Ag/WTiB bilayers, there is detectable silver in low content. In addition, in both variants, trace amounts of elements such as hydrogen, carbon, nitrogen and oxygen can be observed, resulting from standard impurities in the sputtered target and generated during sample transfer. The argon content, on the other hand, results from its use as a working gas during the deposition process. The lower argon content in the Ag/$W_{0.76}Ti_{0.24}B_{2.5}$ sample suggests lower residual stresses [40].

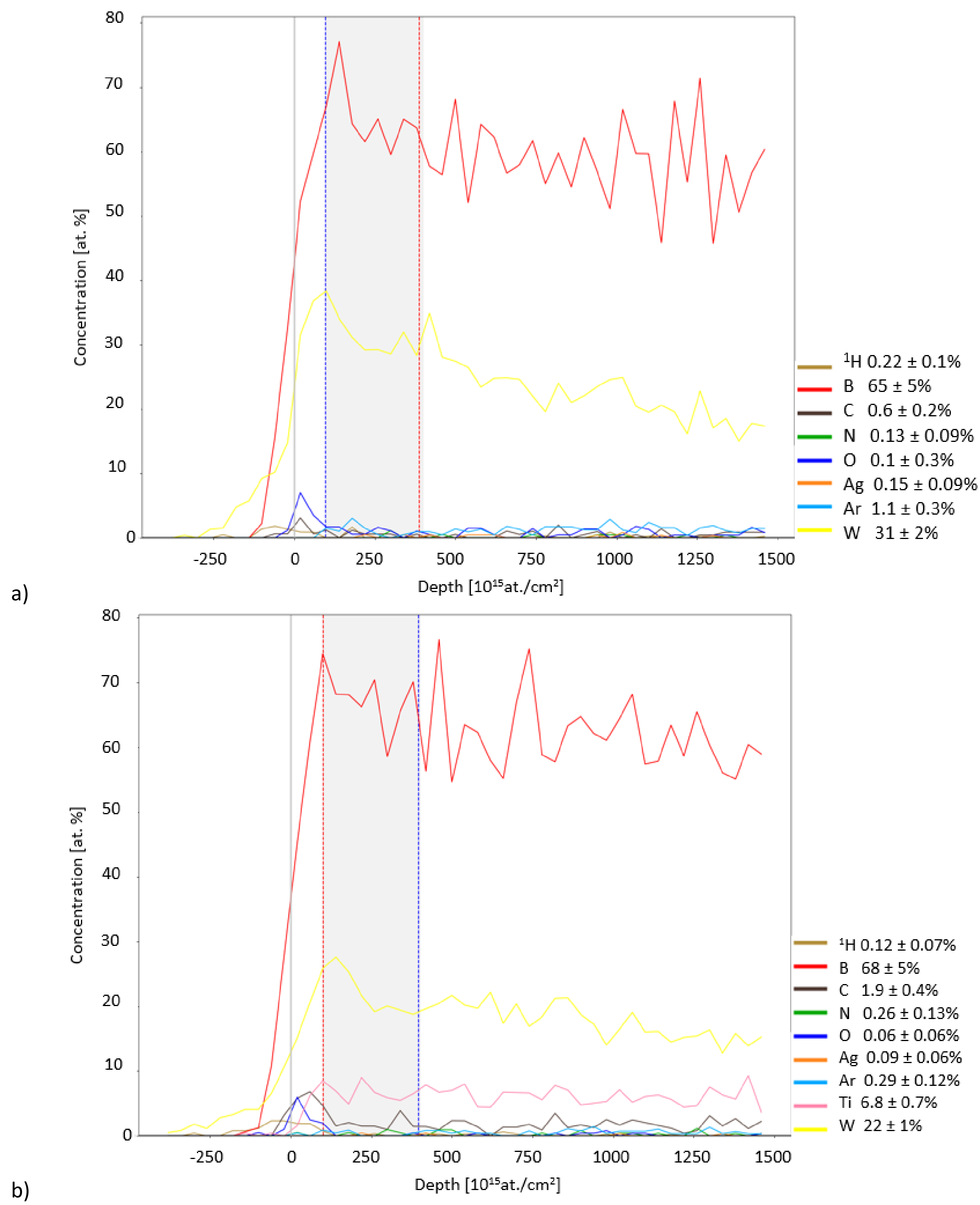

*Figure 6 Chemical composition [at. %] measured on the surface of a) Ag/$WB_{2.5}$ and b) Ag/$W_{0.76}Ti_{0.24}B_{2.5}$.*

### 3.1.2 Surface roughness

The surface roughness of the materials was also measured using an optical profilometer. The average results of the measurements of the deviation of the arithmetic mean from the mean line, along with the measurement error, were determined. For the silver film, the average Ra is 103 ± 10 nm. The relatively high roughness is due to the ablation threshold being exceeded. Above this threshold, ablation efficiency increases because instead of pure evaporation, the plucking of target occurs [41]. The high energy of the laser beam caused the particles to move faster and created so-called droplets on the surface visible in the scanning electron microscope cross-section photo (Figure 2). The droplets formed during laser ablation during layer deposition are usually considered a disadvantage of the PLD method. However, in this particular case, they were used as reinforcement in the composite. The large particle size makes the bilayer rough, and under heavy loads during abrasive wear, the areas with the greatest roughness may break off, and the exposed silver will act as a lubricant.

The roughness of the silver film after deposition with both WB and WTiB decreased, reaching above 70 nm in both cases. According to previous studies, the surface roughness of transition metal boride coatings is only about 1 nm [10]. The high degree of ionisation of the beam in the HiPIMS process means that a significant proportion of the deposited atoms reach the substrate in a directional manner. The ionised atoms fill in local topographical irregularities [42]. Thus, the uniformly deposited boride coatings contributed to the reduced surface development of the pure silver film.

### 3.1.3 XRD

Figure 7 shows X-ray diffraction patterns of the examined coatings. As expected, the laser-produced layer consists of a single-phase material, silver. The graph identifies distinct diffraction peaks with values of 38.09°, 44.32°, and 64.62°, corresponding to the following crystallographic planes: (111), (200), and (202). The atoms are packed as densely as possible, creating a symmetrical, repeating structure. The silver film crystallizes in a face-centered cubic (FCC) structure, corresponding to the Fm-3m space group. The obtained lattice parameter (a ≈ 4.08 Å) is consistent with literature data for silver. The formation of the crystallographic structure correlates well with the TEM image (Figure 3), which clearly shows interference fringes.

In turn, after covering the silver layer with tungsten borides and tungsten doped titanium borides, in both cases the main peak belonging to silver is still visible. Due to the formation of thin films, the X-ray radiation penetrated deeper into the material. A wide hump was visible in the $Ag/WB_{2.5}$ and Ag/WTiB composite layers; both of these materials exhibit amorphous properties, which is also confirmed by the HR – TEM image (Figure 3). However, in the case of the $Ag/WB_{2.5}$ coating, a very small peak corresponding to the $\alpha WB_2$ phase with a hexagonal P6/mmm crystal structure can also be seen in the vicinity of 28°. This is one of the most common hexagonal structures, formed by a metal component from group IV to VI, in this case tungsten. The closely packed metal layers are separated by flat boron layers [43]. The occupation of antibonding states by group VI elements leads to a decrease in the electronic stability of the $AlB_2$ - type structure. Specifically, for these elements, the Fermi level is shifted into the antibonding region of the Density of States (DOS), located above the characteristic pseudogap that separates bonding and antibonding states. Furthermore, vacancies formed in both metallic and boron sites act to stabilize the structure by lowering the Fermi level back towards the pseudogap, thereby reducing the electronic density of states at $E_F$ and minimizing the total energy [44]. Due to the residual stresses in the layer, a shift towards smaller angles occurred. As a result of the lattice transformation, the interplanar distances increased and the lattice parameter c also increased to 3.27 Å.

It is worth noting that in previous studies conducted for the same deposition parameters, the WTiB film had a crystalline structure [10]. However, in the present study, a lack of crystalline order was observed, which may indicate amorphisation of the film. There is a possibility that the presence of significant amounts of alloying additives, such as Ag and Ti, may disrupt the regularity of the crystal lattice. A reduction in the number of vacancies responsible for stabilising the metastable $WB_2$ phase may promote phase transformation. As a result, the $WB_2$ system may tend to form a second, more stable phase, leading to the loss of the original crystallinity of the coating.

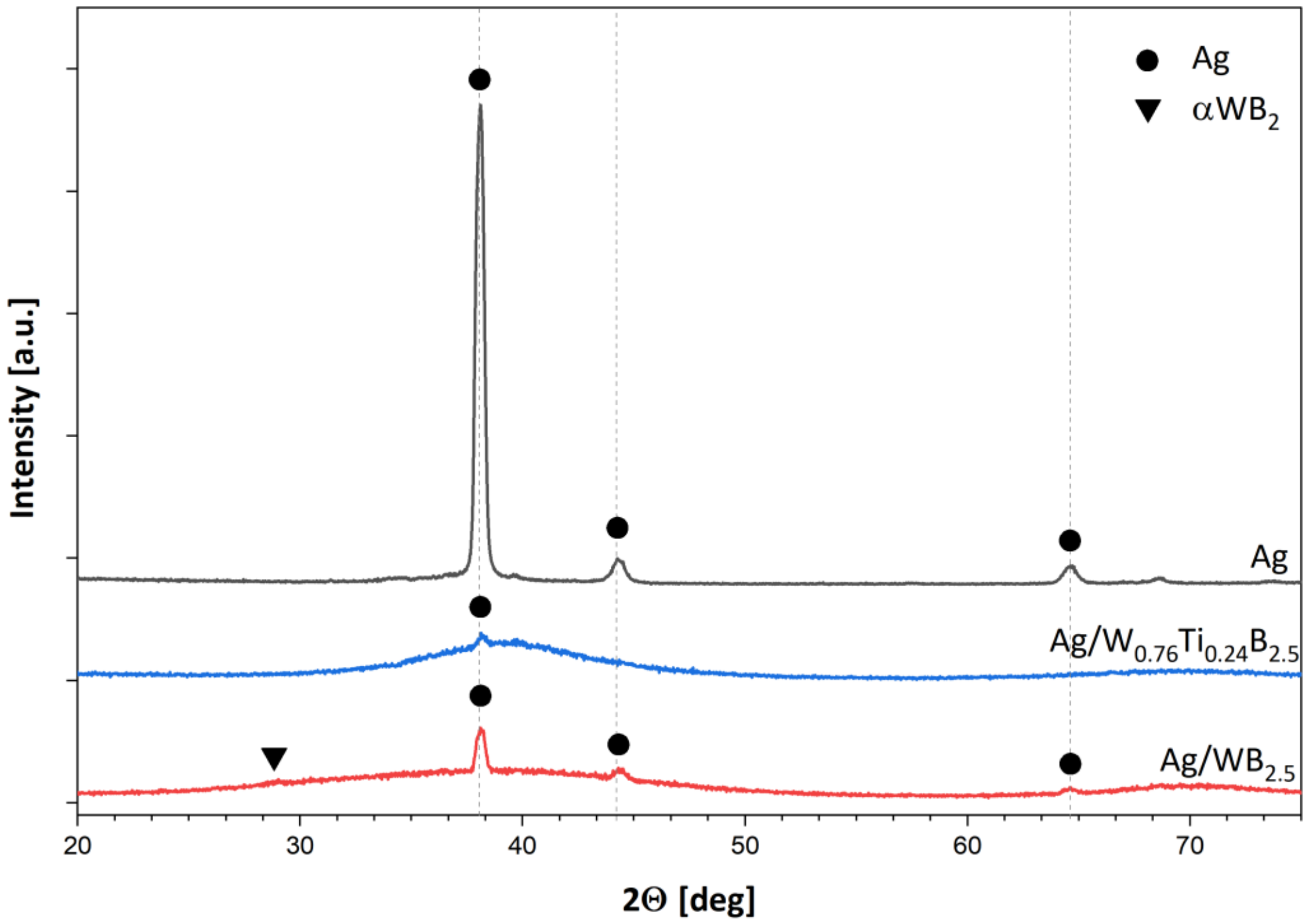

*Figure 7 XRD spectra of Ag, Ag/$WB_{2.5}$ and Ag/$W_{0.76}Ti_{0.24}B_{2.5}$.*

### 3.1.4 XPS

To determine the chemical state of the deposited coatings, XPS analysis was carried out (Figure 8). High-resolution spectral analysis revealed the presence of characteristic core levels of a) Ag 3d, b) B 1s, c) W 4f and d) Ti 2p. In all the analysed coatings, the presence of metallic silver (Ag–Ag) is observed, as well as components associated with the presence of silver oxides (Ag - O - C, Ag - O), indicating partial oxidation of the surface coating. The Ag 3d spectra show characteristic peaks at binding energies of approximately 368.4 eV (Ag $3d_{5/2}$) and 374.4 eV (Ag $3d_{3/2}$), corresponding to spin - orbit splitting. The positions of these peaks are consistent with literature data and indicate the dominant presence of metallic silver, alongside the presence of its oxidised forms [45]. The formation of silver oxides on the surface may have potential applications in the future as additional antibacterial layers.

For coatings containing boron and tungsten, the B 1s and W 4f spectra confirm the presence of W - B bonds, characteristic of boron phases. The binding energy of the W $4f_{7/2}$ peak is approx. 32 eV, whilst for B 1s it is approx. 189 eV, which is consistent with literature data for tungsten borides [46]. Compared to pure tungsten (approx. 31.4 eV), a shift towards higher binding energies is observed, which may indicate a change in electron density around the tungsten atoms, associated with interaction with boron and the presence of oxygen. In the W 4f spectra, visible components corresponding to tungsten oxides confirm partial surface oxidation. In the Ag/$W_{0.76}Ti_{0.24}B_{2.5}$ coating, a Ti 2p signal was additionally recorded. Analysis of its components indicates both the presence of Ti - B bonds and the presence of titanium oxides (Ti - O). The observed shift relative to the value for pure titanium (454 eV) towards higher binding energies is typical of oxidised states of titanium.

Oxygen and carbon are also observed in all spectra, which is typical of XPS analysis and stems from the surface-specific nature of this method. It should be emphasised that XPS analyses only the surface layer; therefore, the detected presence of oxides may be related to contact with air.

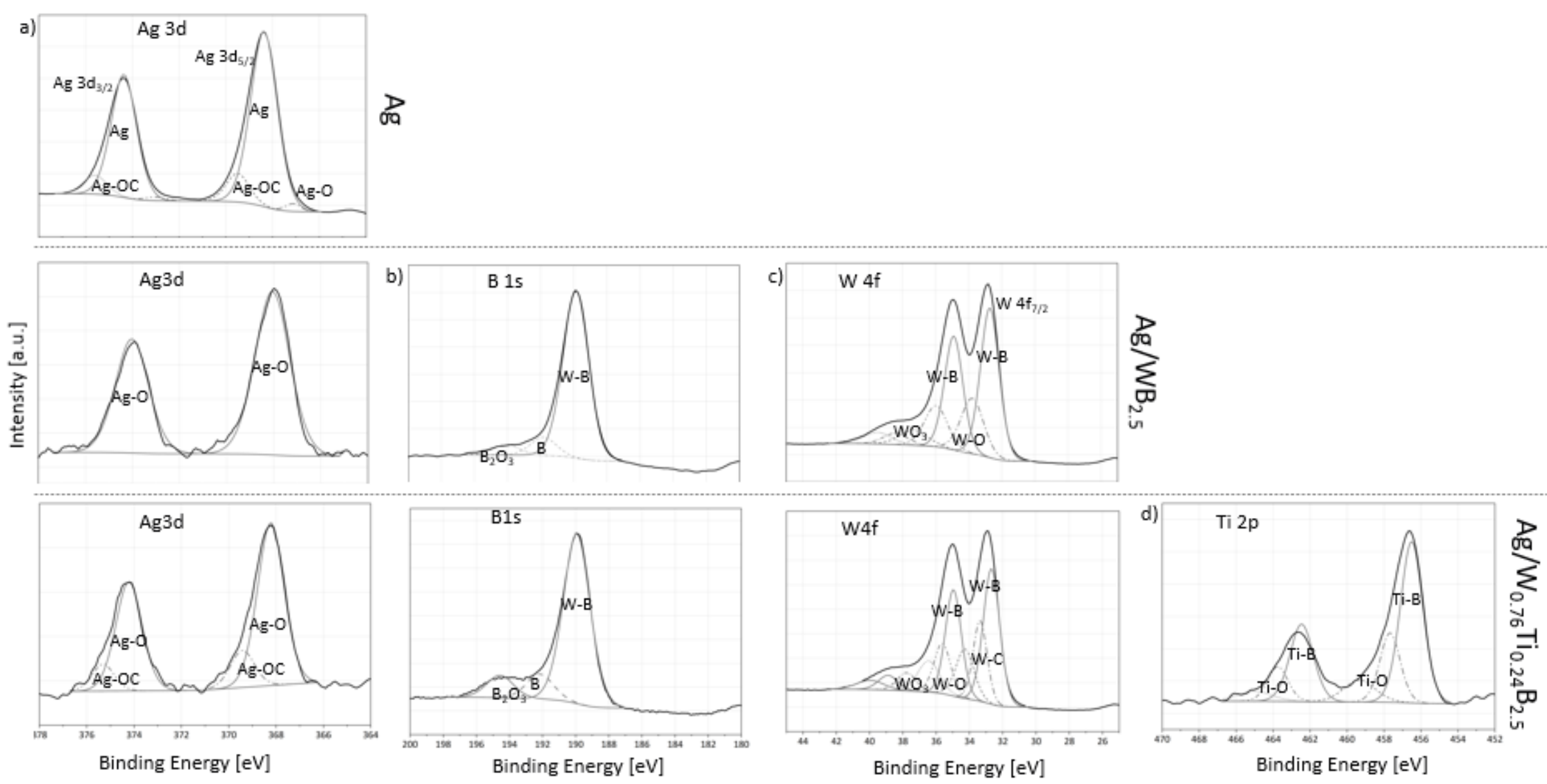

*Figure 8 XPS analysis of Ag, Ag/$WB_{2.5}$ and Ag/$W_{0.76}Ti_{0.24}B_{2.5}$ high-resolution spectra of selected core levels: a) Ag 3d, b) B 1s, c) W 4f, and d) Ti 2p.*

## 3.2 Mechanical and tribological properties

### 3.2.1 Hardness

The mechanical properties of the produced materials were then investigated. Figure 9 a) shows the relationship between penetration depth and hardness. To assess the stiffness of the material, Figure 9 b) shows Young's modulus as a function of penetration depth. In the case of the Ag coating, the hardness stabilises at depths above ~150 nm and is approximately 2.5 GPa; however, this region is likely influenced by the substrate. Therefore, the hardness of the coating should primarily be considered in the lower range of indentation depth, where the influence of the substrate is minimised. The maximum hardness of the silver layer for a depth not exceeding 100 nm is approx. 1.7 GPa. These values are typical for crystalline metallic materials, in which plastic deformation is mainly caused by the movement of dislocations in the crystal lattice.

By coating the WB and WTiB silver films, bilayers were created with maximum hardness of 22.06 ± 3.03 GPa and 25.73 ± 1.05 respectively. According to the results of XRD analysis, the obtained bilayers exhibit an amorphous structure, which results in a reduction in hardness compared to boron layers alone [10]. The lack of long-range crystalline order suppresses conventional plastic deformation mediated by dislocations, which is characteristic of crystalline metals. Instead, deformation occurs through local atomic rearrangements, which typically require higher stresses. It is worth noting that the presence of the silver film contributed to an increased dispersion of hardness results, which is associated with greater surface roughness of the tested samples.

The lowest Young's modulus was recorded for pure silver and was around 140 GPa for penetration depths below 100 nm. The relatively low Young's modulus compared to other metals makes the material quite elastic. As a result of coating the silver with WB and WTiB, the stiffness of the material increased significantly. The Young's modulus of the Ag/WB composite was as high as 172.2 ± 14.1 GPa, whilst that of the Ag/WTiB composite was 244.5 ± 36.01 GPa. However, for materials with such high hardness, the values obtained are not particularly high compared to commonly known carbides [47].

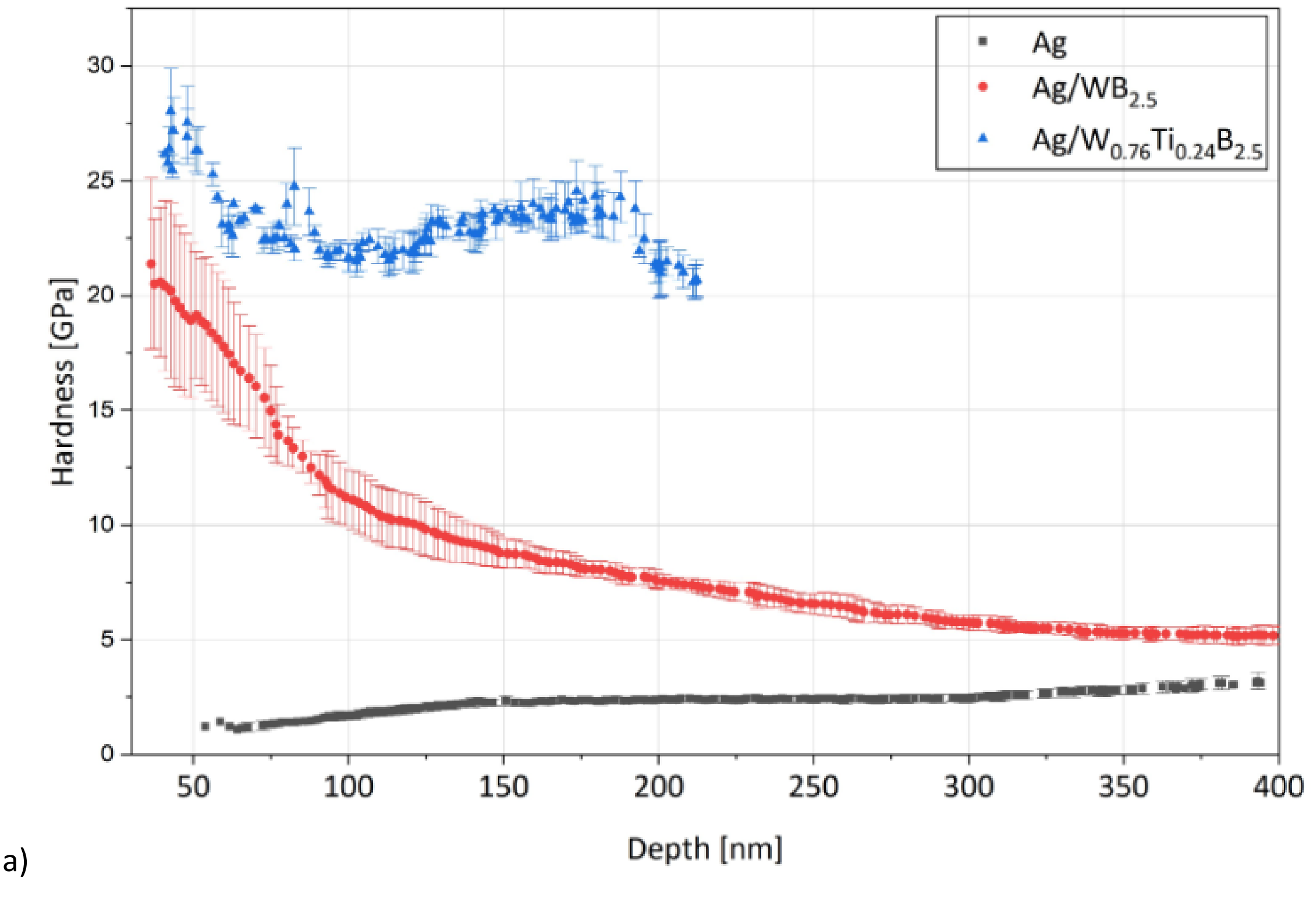

a)

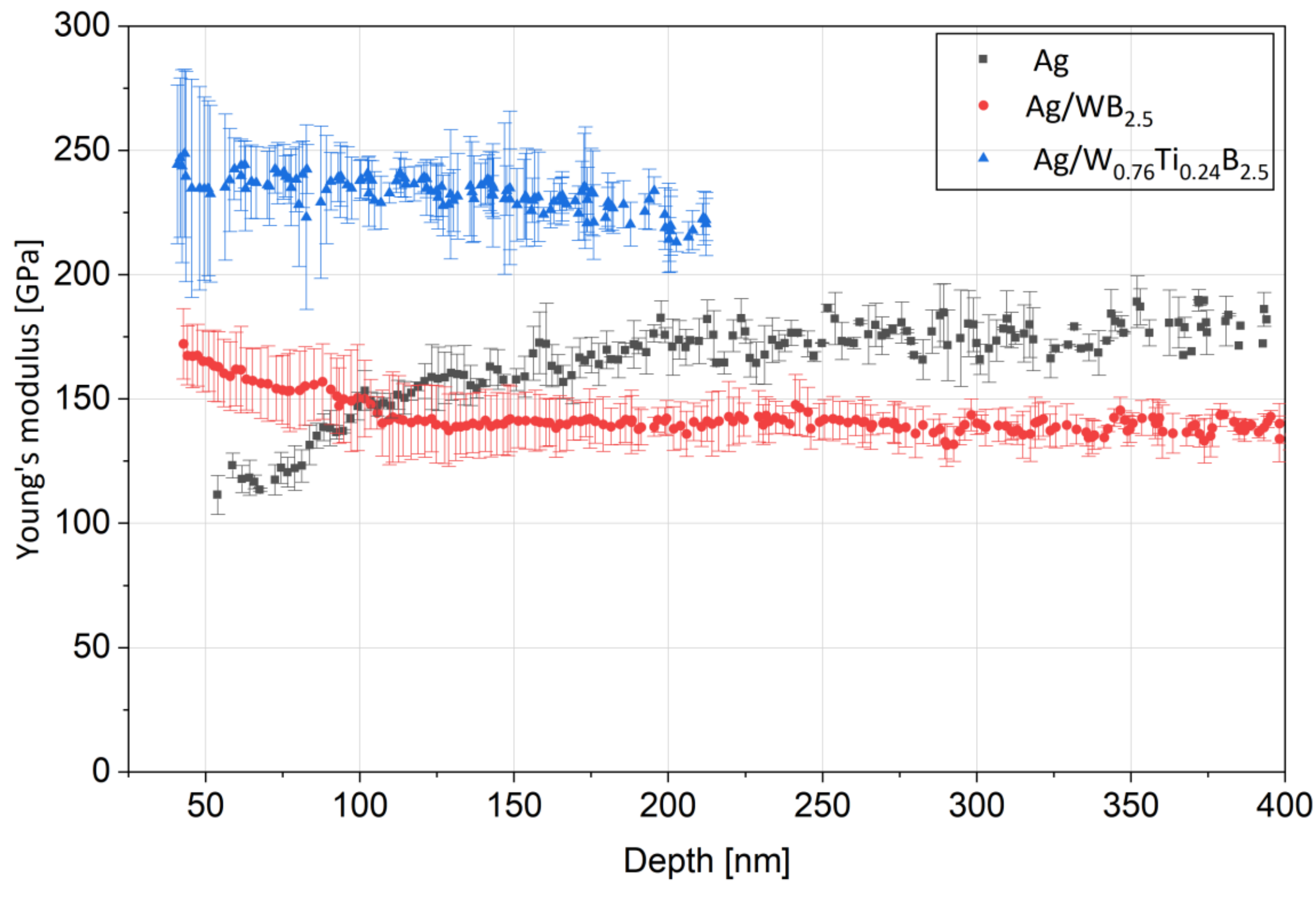

b)

*Figure 9 a) Hardness and b) elastic modulus of Ag, Ag/$WB_{2.5}$ and Ag/$W_{0.76}Ti_{0.24}B_{2.5}$ coatings as a function of indentation depth, determined using the Continuous Stiffness Measurement (CSM) method.*

### 3.2.2 Fracture Toughness

In order to assess the fracture toughness, microhardness impressions were made using a cube corner indenter. Figure 10 shows selected images of the imprints by SEM. Measurements were taken at loads of 200 mN and 300 mN (Figure 10). The test was carried out for two loads in order to verify that the results obtained were not influenced by the substrate used. The average results of indentation resistance measurements using the nanoindentation method with a cube-corner indenter were also calculated. In the case of the film composed of pure silver, no typical radial cracking was observed (Figure 10 a)). However, a large accumulation of material at the edges of the imprint can be seen in the image, indicating a high plasticity of the material. Part of the coating piled up, which is characteristic of soft metals. In this case, the indentation method used to determine $K_C$ does not provide sufficient reliability of results. For this reason, for pure silver film, only a qualitative analysis of the morphology of the indentations was performed.

After the application of the tungsten boride coating, a series of cracks can be seen in the photo as a result of the indentation (Figure 10 b)). This is typical behaviour of a brittle material. There are relatively long cracks, approximately 35 µm, extending radially from the corners of the cube corner imprint. The WB film not only has high hardness, but also the very hight Young's modulus. As a result of local tensile stresses generated by the indenter, the energy is not dissipated by plastic deformation, but cracks and chips are formed. In contrast, almost no cracks were observed around the imprints after imprinting in the Ag/WTiB bilayer (Figure 10 c)). The addition of titanium significantly improved the resistance to brittle fracture; the coating is cohesive. The metal addition to the layer significantly increased the plasticity of the material, reduced stresses and improved adhesion. The XPS analysis revealed that titanium in the studied coatings is present not only in the form of titanium oxides but also in chemical states corresponding to Ti - B bonding (Figure 8). This indicates the formation of a multicomponent, complex structure containing different local bonding such as W - B and Ti - B. The literature indicates that the presence of titanium boride can positively influence the resistance of materials to brittle fracture. Microstructural or chemical inhomogeneities can act as barriers to crack propagation [48, 49]. In the case of the Ag/WB film, in which numerous and relatively long cracks were observed, the $K_C$ values obtained were approximately 0.1 MPa√m. For the Ag/WTiB layer, characterized by significantly shorter cracks, the $K_C$ values were significantly higher, at 5.84 ± 0.09 MPa√m and 4.8 ± 0.51 MPa√m for loads of 200 mN and 300 mN, respectively. At lower loads, shorter cracks are observed, which leads to seemingly higher $K_C$ values. This phenomenon is related, among other things, to the indentation size effect and a greater proportion of plastic deformation in a small volume of material. In both analyzed coatings, the results obtained are consistent with the observed morphology of the indentations - shorter cracks indicate higher resistance of the material to brittle fracture.

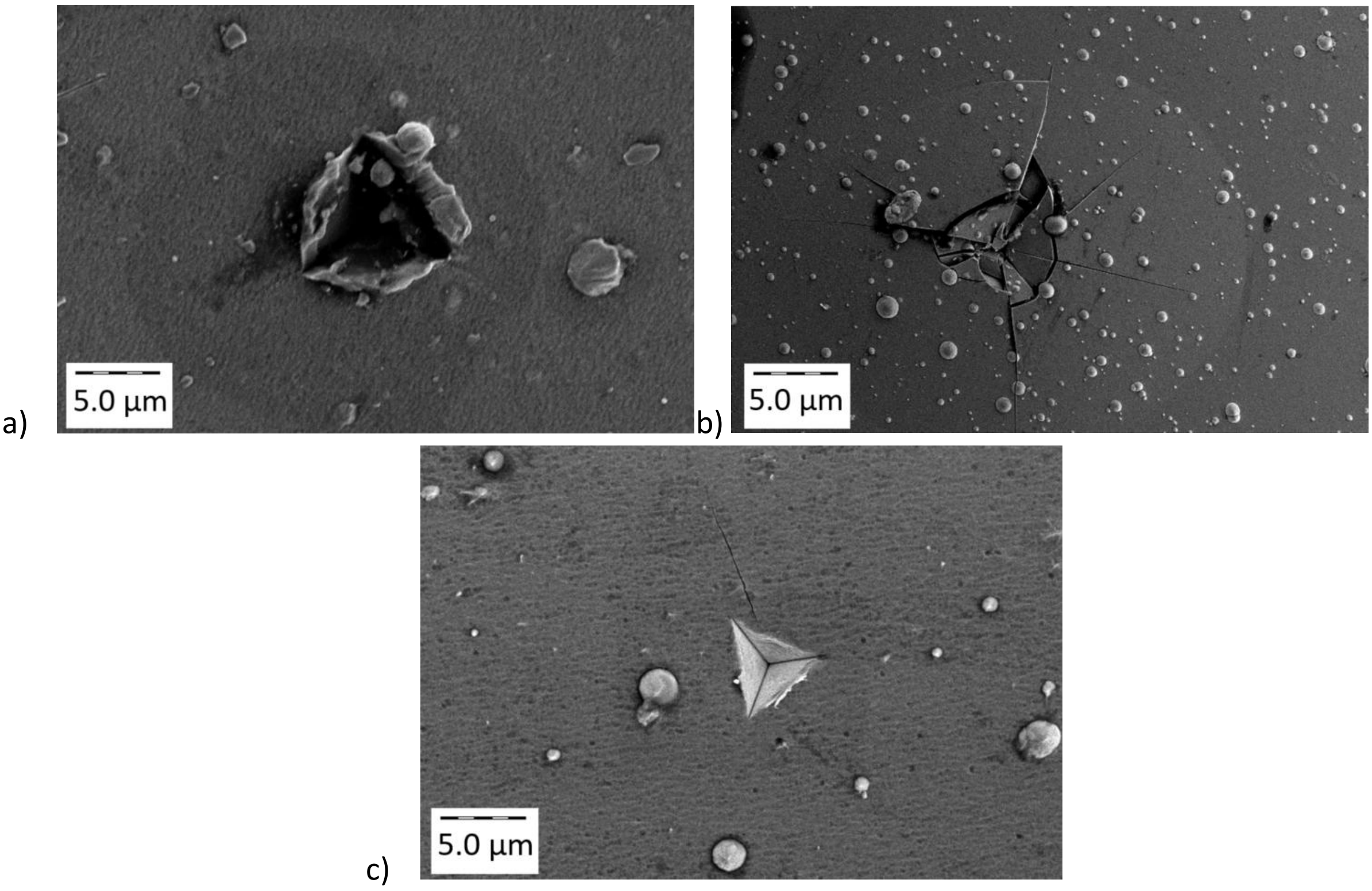

*Figure 10 Indents (cube corner indenter) SEM images for coatings of a) Ag, b) $Ag/WB_{2.5}$, c) $Ag/W_{0.76}Ti_{0.24}B_{2.5}$; load: 300 mN.*

### 3.2.3 Scratch test

The adhesion of the coatings to the substrate was assessed using adhesion tests: 'scratch-test'. The scratch length for all tested coatings was 3 mm. Figure 11 shows images of the resulting scratches taken with a light microscope. The nature of the damage varies depending on the sample tested. In the case of pure silver, the film undergoes so-called plastic flow. The coating does not fracture, but streaks form along the indenter action (Figure 11 a)). This phenomenon is particularly visible at higher approximations (Figure 12 a)). Additionally, it can be observed that part of the film undergoes plastic pile-up, with material accumulating on the sides of the forming scratch. This is a typical feature among relatively soft materials. This observation is confirmed by EDS analysis of the scratch. Table 3 presents the chemical composition measured on the surface of the materials at the marked points shown in Figure 12. Within the visible scratch on the silver-coated surface, signals originating from the substrate (approximately 20 at. % Si) were detected in addition to silver, indicating partial exposure or penetration of the substrate. In contrast, regions located at the edge of the scratch show an accumulation of nearly pure silver. When the $Ag/WB_{2.5}$ bilayer was scratched, the material was completely delaminated. Already after passing about 0.7 mm in length, the coating was completely destroyed. Figure 12 b) shows fragments of the brittle, delaminated layer. Analysis of the chemical composition has shown that the dark areas represent the base of the material, so the film does not have adequate adhesion. This is also associated with a higher Young's modulus of the $WB_{2.5}$ film. The material is unable to deform as a result of the generation of very high shear stresses and immediately begins to crack. In areas where the coating is still visible, the chemical composition differs significantly from the results of the ToF-ERDA analysis (Figure 6). Due to the low X-ray yield of boron, quantitative determination of B by EDS is subject to significant uncertainty. Therefore, the ToF-ERDA results are considered more representative of the actual coating composition. The addition of Ti to the bilayer resulted in a significant change in properties. When approximated with an optical microscope, the crack is almost invisible (Figure 11 c)). However, with a closer approximation, it can be seen that the damage is not in the form of brittle microcracks typical of ceramic materials (Figure 12 c)). The image shows a longitudinal streak along the indenter action, similar to that of a pure metal film. The bilayer is characterised by very good adhesion and has high scratch resistance. The coating has more matched linear thermal expansion coefficient to the silver layer, making it more mechanically stable. Furthermore, no cracks are observed on the interface between silver - boron tungsten doped with titanium (Figure 4 c)). The matrix material thoroughly covers the surface of the reinforcement particles, which increases adhesion.

Figure 13 a) illustrates a graph of the evolution of acoustic emission. The evolution of acoustic emission during crack formation correlates well with microscopic observations. Almost at the very beginning of the indenter passage through the silver film, there was a peak in the acoustic wave emission. The intense signal was associated with damage initiation. Thereafter, the emission value decreased and remained at a constant low level. This result indicates very good scratch resistance of the silver. It has a ductile structure and undergoes stable deformation. In the case of the Ag/$WB_{2.5}$ coating, once the scratch length exceeds 0.7 mm, the emission value increases considerably, undergoing continuous perturbation. The nature of this type of curve is due to poor adhesion and cracking of the coating. In contrast, the acoustic emission evolution curve of the Ag/$W_{0.76}Ti_{0.24}B_{2.5}$ material has a few peaks, which are most likely due to minor damage. A plot of the change in friction coefficient during scratching is also shown (Figure 13). All coatings had relatively low friction coefficients. However, during data recording of the Ag/$WB_{2.5}$ and Ag/$W_{0.76}Ti_{0.24}B_{2.5}$ coatings, there were significantly larger coefficient fluctuations compared to the pure silver film. A lower coefficient of friction means that the surfaces of the materials can move more easily between each other. By knowing information about the tribology of the tested coatings, it is possible to increase the durability of the materials by reducing abrasive wear. The critical load $Lc_1$, i.e. the load at which the first failure of the material films occurs, are respectively:

-1.46 N for the Ag coating,
- 0.3 N for the Ag/$WB_{2.5}$ coating,
- 2.33 N for the Ag/$W_{0.76}Ti_{0.24}B_{2.5}$ coating.

a)
b)
c)

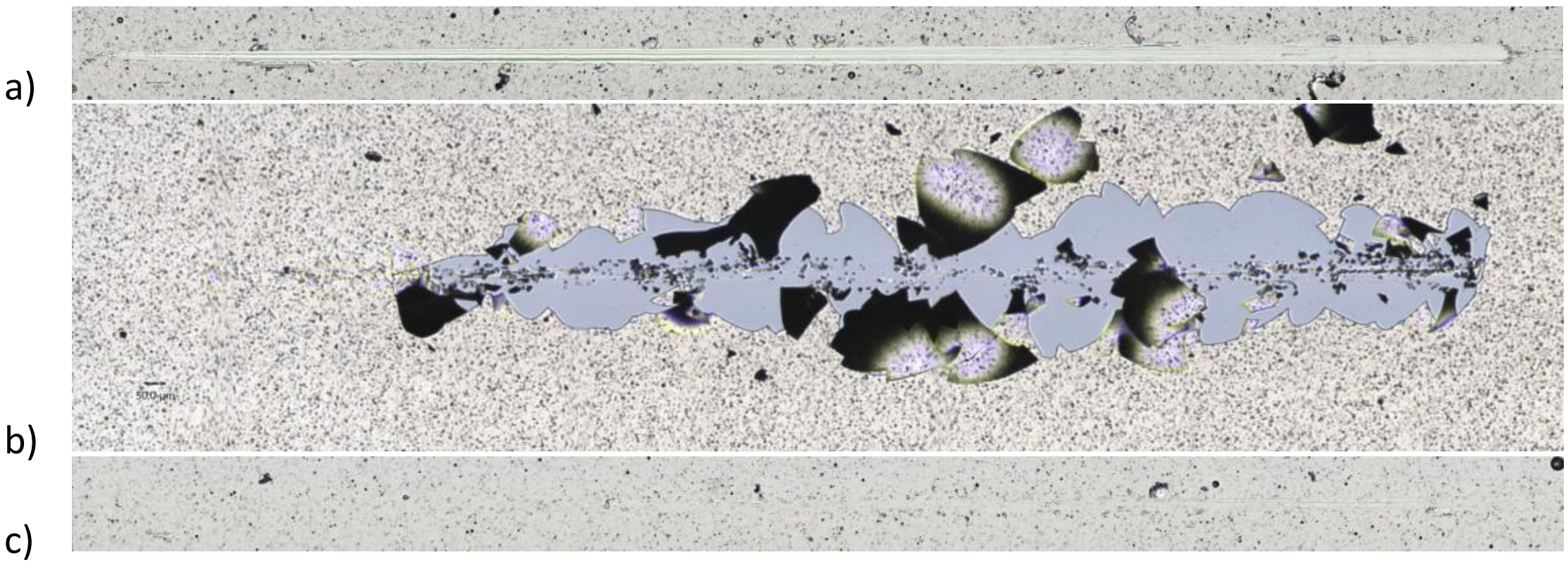

*Figure 11 Light microscope images of the surface morphology of the scratch track for coatings a) Ag, b) Ag/$WB_{2.5}$, c) Ag/$W_{0.76}Ti_{0.24}B_{2.5}$.*

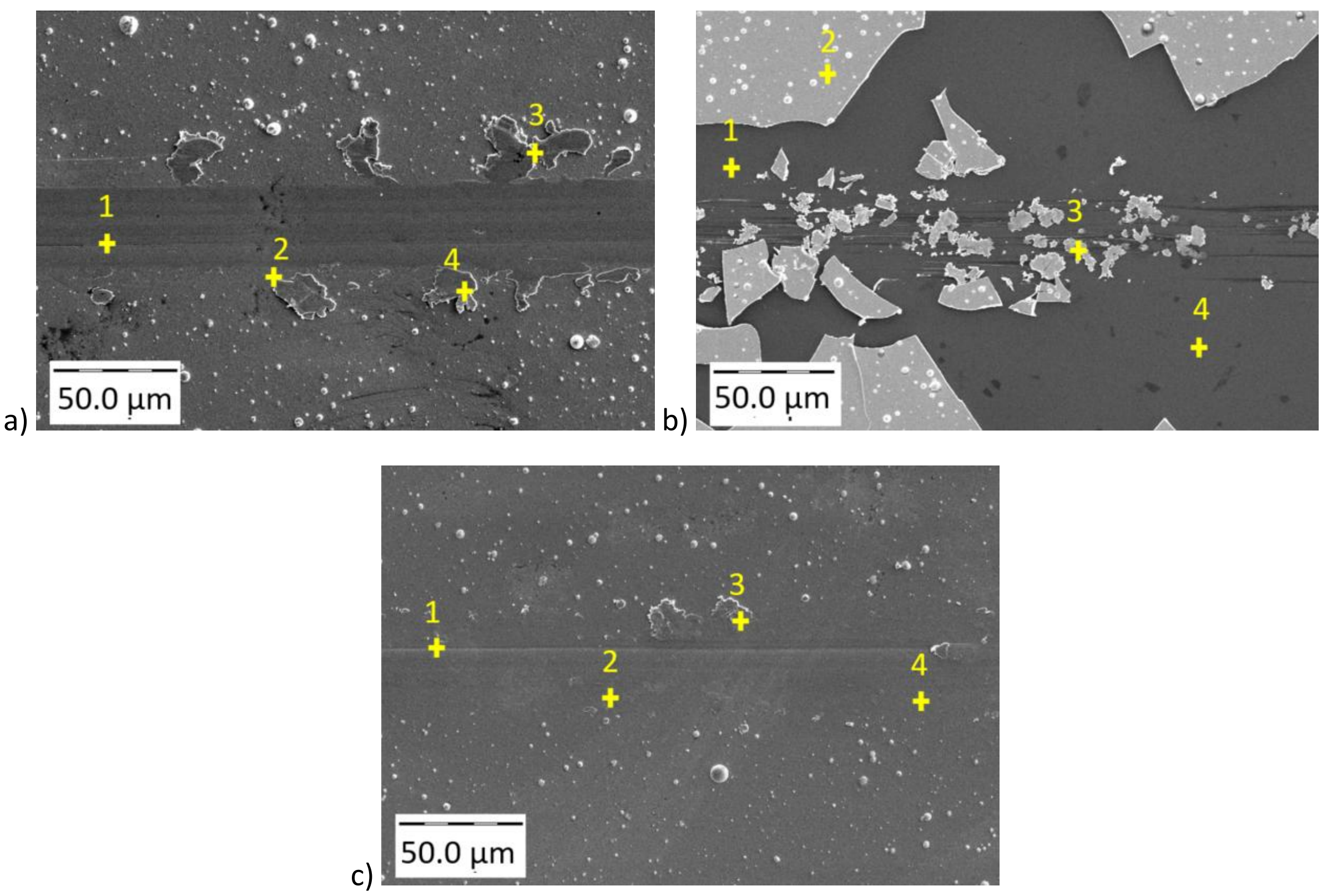

*Figure 12 SEM images of scratch track with marked EDS measurement points for coatings of a) Ag, b) $Ag/WB_{2.5}$, c) $Ag/W_{0.76}Ti_{0.24}B_{2.5}$.*

*Table 3 Chemical composition [at. %] measured on the surface of: Ag, $Ag/WB_{2.5}$, $Ag/W_{0.76}Ti_{0.24}B_{2.5}$; scratch test.*

| Materials | Point number | Atomic content [%] | | | | | |
|---|---|---|---|---|---|---|---|
| | | Ag | Si | W | B | Ti | O |
| Ag | 1 | 76.2 | 23.8 | | | | |
| | 2 | 100 | | | | | |
| | 3 | 99.9 | 0.1 | | | | |
| | 4 | 99.6 | 0.4 | | | | |
| $Ag/WB_{2.5}$ | 1 | | 100 | | | | |
| | 2 | | | 42.4 | 52.2 | | 5.3 |
| | 3 | | 100 | | | | |
| | 4 | 8.4 | 89.2 | | | | 2.3 |
| $Ag/W_{0.76}Ti_{0.24}B_{2.5}$ | 1 | 0.4 | | 45.9 | 36.7 | 17.1 | |
| | 2 | 4.8 | | 45 | 33.4 | 16.8 | |
| | 3 | 0.7 | | 55 | 30.5 | 13.8 | |
| | 4 | 0.6 | | 38.1 | 45 | 16.3 | |

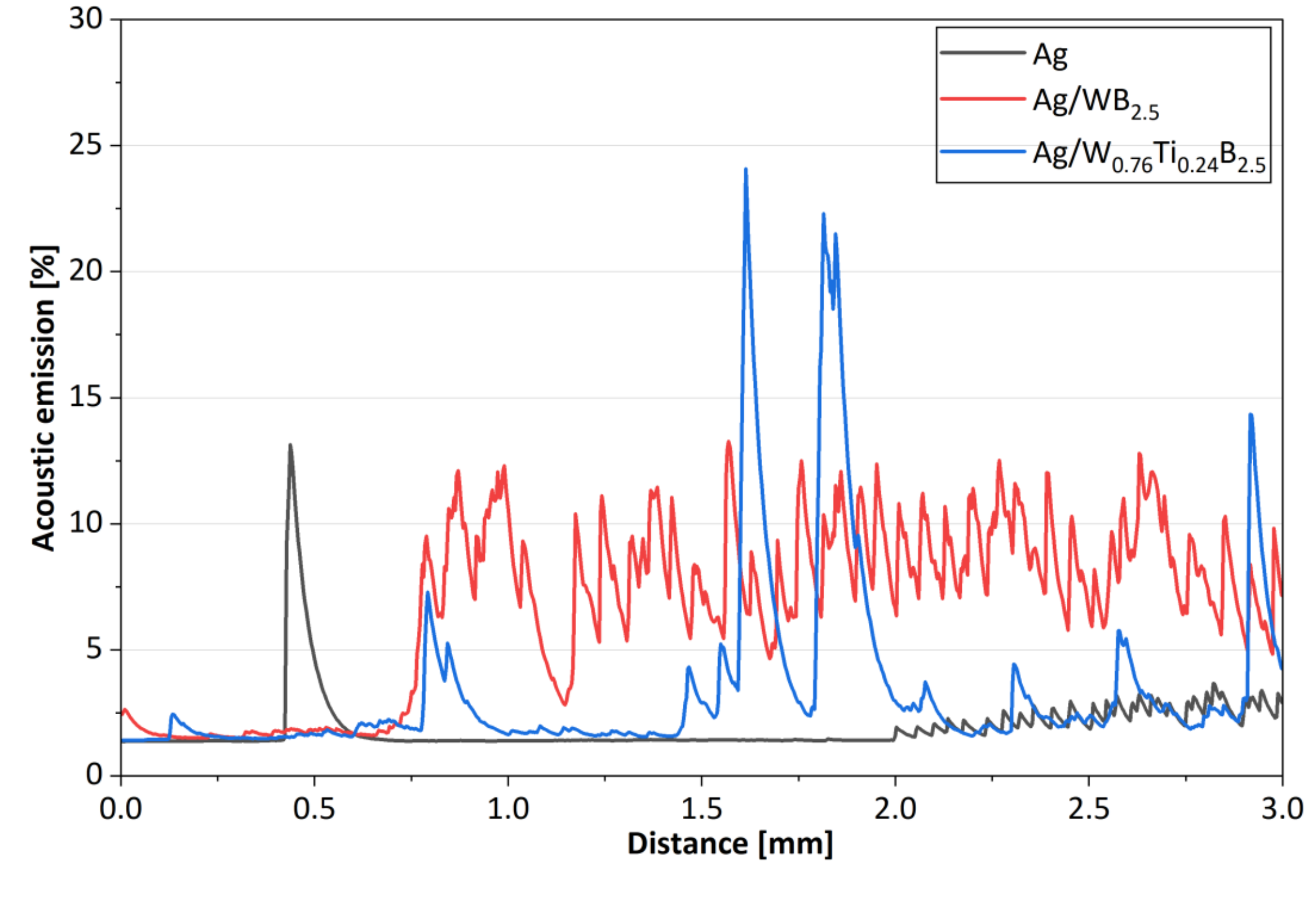

a)

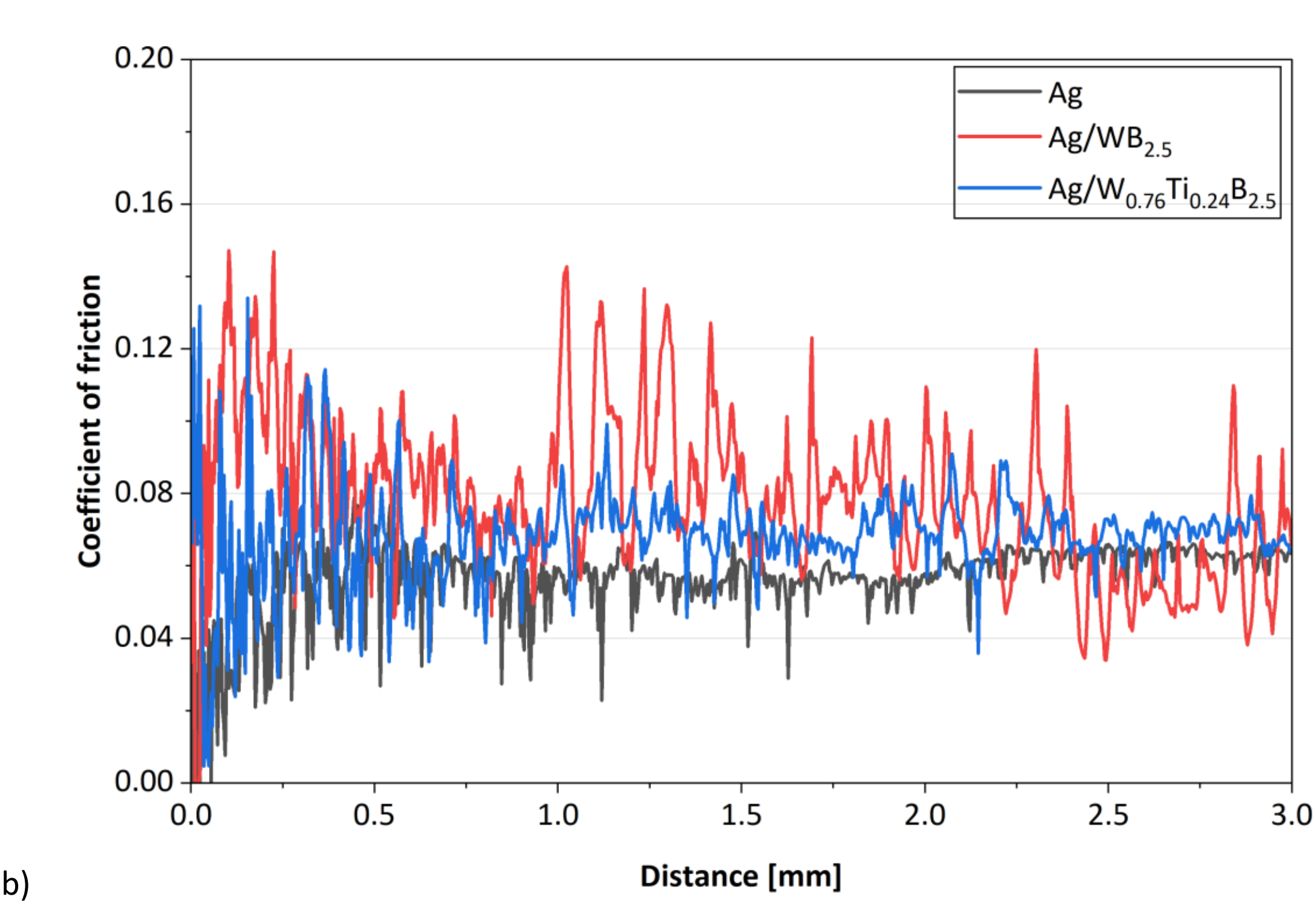

b)

*Figure 13 a) Evolution of the acoustic emission, b) evolution of the coefficient of friction.*

### 3.2.4 Wear resistance

Figure 14 shows optical microscope images of the wear tracks after the abrasion process in reciprocating motion. Pure silver shows very good tribological properties. Due to its relatively low hardness, it easily undergoes plastic deformation. No cracks can be seen in the picture, the wear trace is uniform, and the so-called lubrication effect was created [20]. Along the direction of movement of the alumina ball, the material smeared (Figure 14 a)). In contrast, Figure 14 b), c) shows the wear trace of the Ag/WB and Ag/WTiB bilayers. In both cases, darker areas are visible, resulting from long-term sliding of the antibody across a dry, unlubricated surface. These areas correspond to wear zones where partial degradation of the layer has occurred. It is noteworthy that the bilayers have, for the most part, not become detached from the substrate, indicating their good resistance to adhesive wear. After performing EDS analysis for the Ag/$W_{0.76}Ti_{0.24}B_{2.5}$ coating, the chemical composition within the wear mark remains similar to the initial composition of the coating. The distribution of elements on the surface shows no significant changes; it still consists primarily of boron, tungsten, titanium and a small amount of silver. No clear signs of intensive material transfer or the formation of new reaction products is observed.

Figure 15, on the other hand, shows a plot of the evaluation of the coefficient of friction. A typical friction curve as a function of the number of movement cycles can be divided into several phases. For all layers tested, the friction coefficient increases with increasing distance. The actual contact area increases with the number of movement cycles, which leads to an increase in adhesive friction. The increase in the friction coefficient documents the dominant contribution of adhesive friction [19]. For the film composed of pure silver, the changes in the friction coefficient are the smallest relative to the other coatings. The maximum value of the coefficient is only 0.29. During sliding contact, tiny silver particles are crushed and sheared, acting like “lubricant” in a solid body. Wear particles, as third bodies, play an important role in contact friction and can reduce friction and wear by creating slippery materials [50]. The recorded values correlate very well with the wear trace photo (Figure 14 a)). The good behaviour of the sample containing the silver film is confirmed by the low value of the wear coefficient.

After the application of ceramic coatings WB and WTiB, the coefficient of friction increased, reaching maximum values of 0.45 and 0.84, respectively. It is most likely that the addition of titanium in the ceramic coating caused greater adhesion to the counterexample, thus increasing the friction between the contacting surfaces [51]. The greatest effective wear depth was recorded for the Ag/WB layer, amounting to 1.27 ± 0.05 (Table 4). Intensive wear led to almost complete abrasion of the coating down to the substrate, which is consistent with the reduced friction coefficient. As a result, the corundum ball slid directly on the silicon.

*Table 4 Material loss per unit area after wear testing.*

| Sample | Effective wear depth [µm] |
|---|---|
| Ag | 0.68 ± 0.11 |
| Ag/$WB_{2.5}$ | 1.27 ± 0.05 |
| Ag/ $W_{0.76}Ti_{0.24}B_{2.5}$ | 0.62 ± 0.01 |

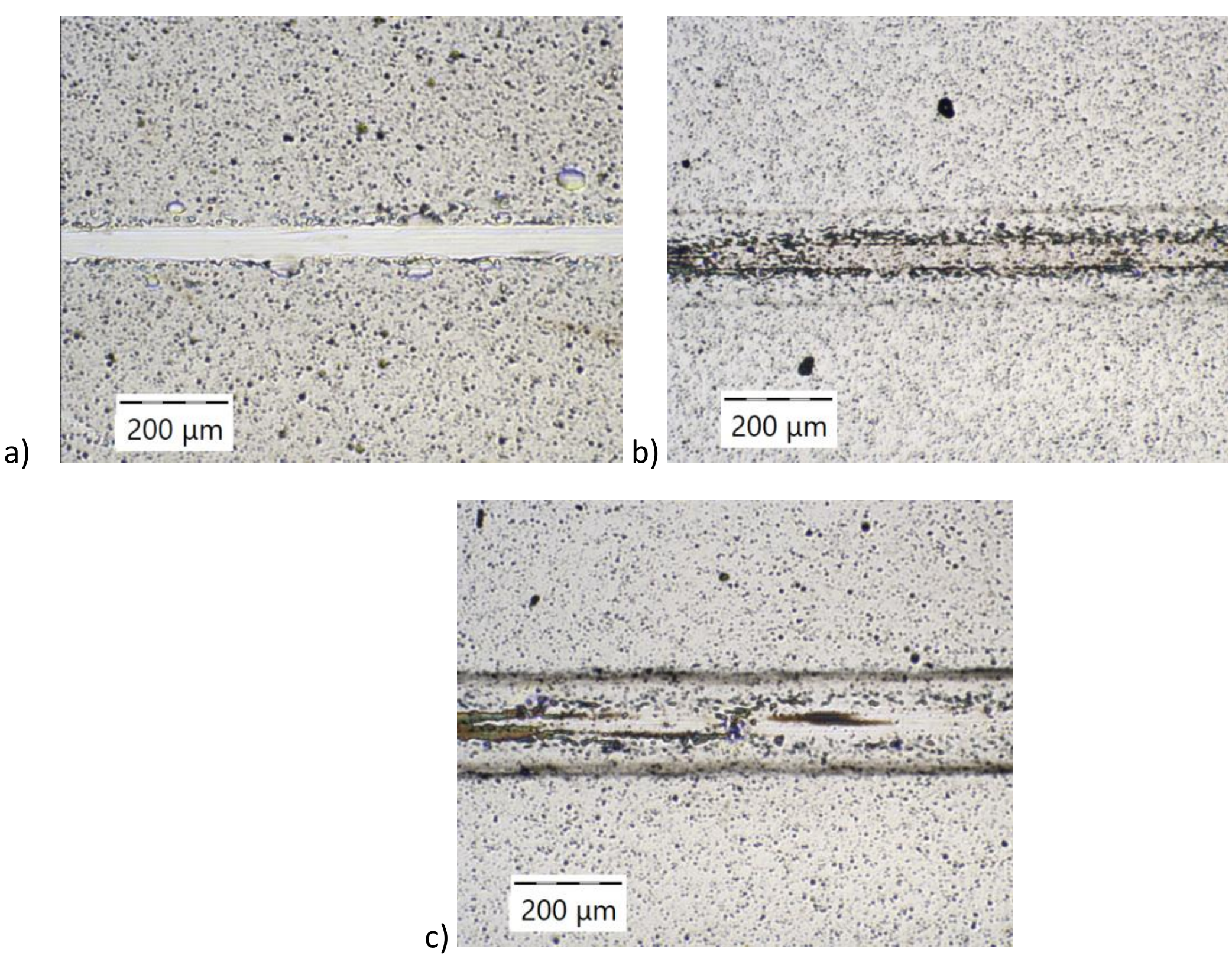

*Figure 14 Images of the wear track obtained during the abrasion in reciprocating motion in coating a) Ag, b) $Ag/WB_{2.5}$, c) $Ag/W_{0.76}Ti_{0.24}B_{2.5}$.*

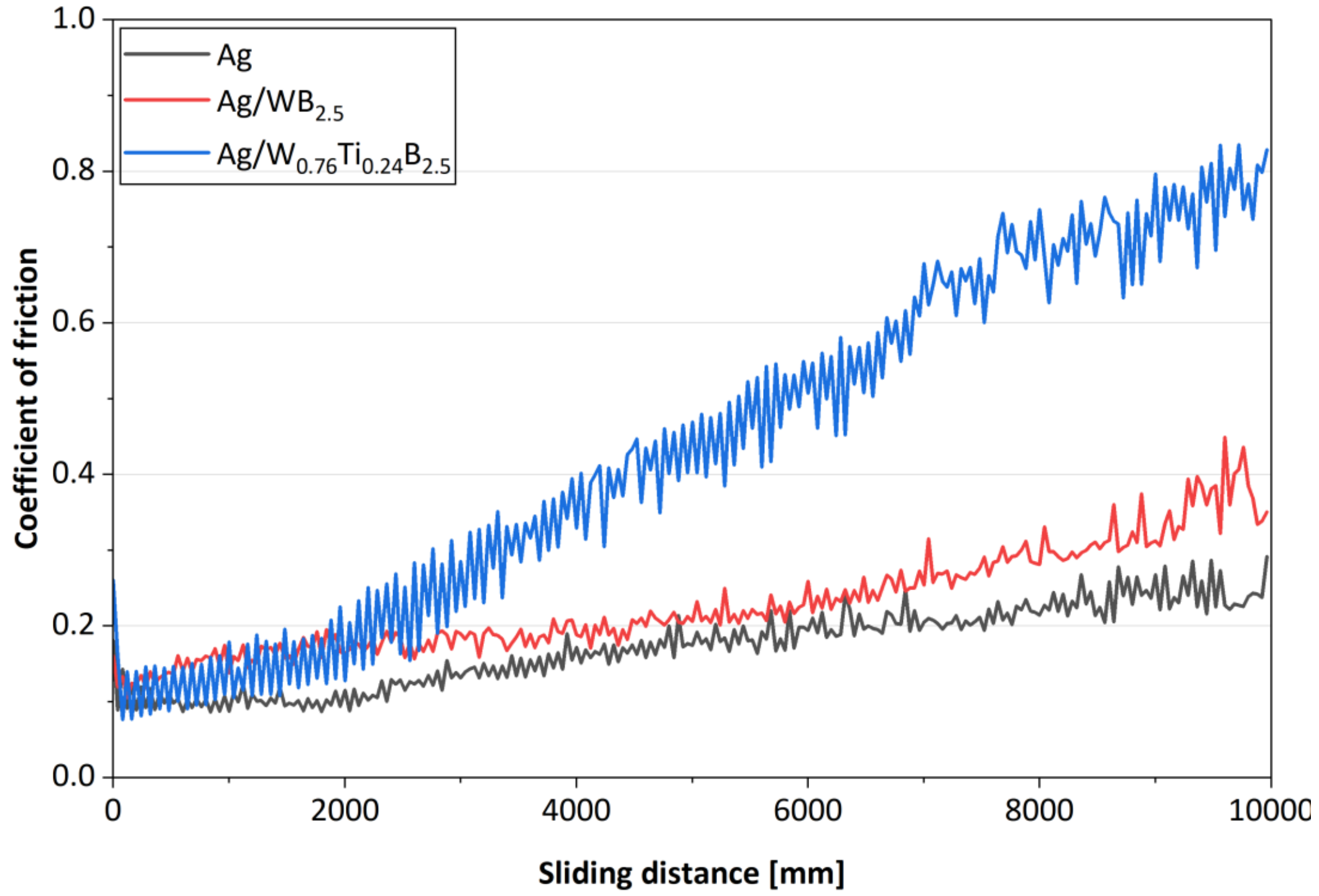

*Figure 15 Evolution of the Coefficient of Friction for coating Ag, $Ag/WB_{2.5}$, $Ag/W_{0.76}Ti_{0.24}B_{2.5}$.*

### 3.2.5 Tribotouch

The perception of a material's quality depends primarily on its appearance, which is why in this study tests were also performed for resistance to manual abrasion, the so-called hand abrasion simulation. This is an innovative and increasingly used test for assessing the durability of coatings under conditions simulating real, everyday use. Due to the inferior adhesion of the Ag/$WB_{2.5}$ bilayer, the tests were carried out on two samples, containing an Ag film and an Ag/WTiB film as well. After 10000 cycles, no macroscopic changes were found in either the film containing pure silver or in the boride-coated coating, indicating their very good abrasion resistance. After 11000 cycles for silver, the sample was subjected to microscopic observations (Figure 16 a)). The surface of the material had changed significantly compared to its initial state (Figure 2 a)). As a result of continuous tapping with a simulated finger, the layer no longer shows so-called droplets, but a series of parallel cracks. Due to the relatively low Young's modulus of the silver, the material did not splinter but deformed plastically. The chemical scaling of the surface after wiping was also investigated. Primarily silver is still detectable on the surface (95.6% at.). The analysis also revealed the presence of oxygen (3.4% at.)., indicating that the film had undergone oxidation, as well as silicon (1% at.), originating from the substrate. In contrast, after 36000 abrasion cycles of the Ag/WTiB sample, droplets are still visible on the surface of the material (Figure 16 b)). As a result of the friction of the sample, the surface was not damaged, but the droplets were deformed, along the push-slip force. EDS analysis showed no major changes in the material composition. The coating is still composed of silver (0.6% at.) and transition metal borides (44.2% at – W, 37.6% at – B, 12.3% at – Ti). As a result of the abrasion, the amount of silver increased minimally compared to the initial state and the appearance of oxygen on the surface was detected. It is noteworthy that both the pure silver compound film after 11000 cycles and the tungsten boride-coated silver compound film after 36000 cycles showed no detachment from the substrate. This test showed very good adhesion of the bilayer.

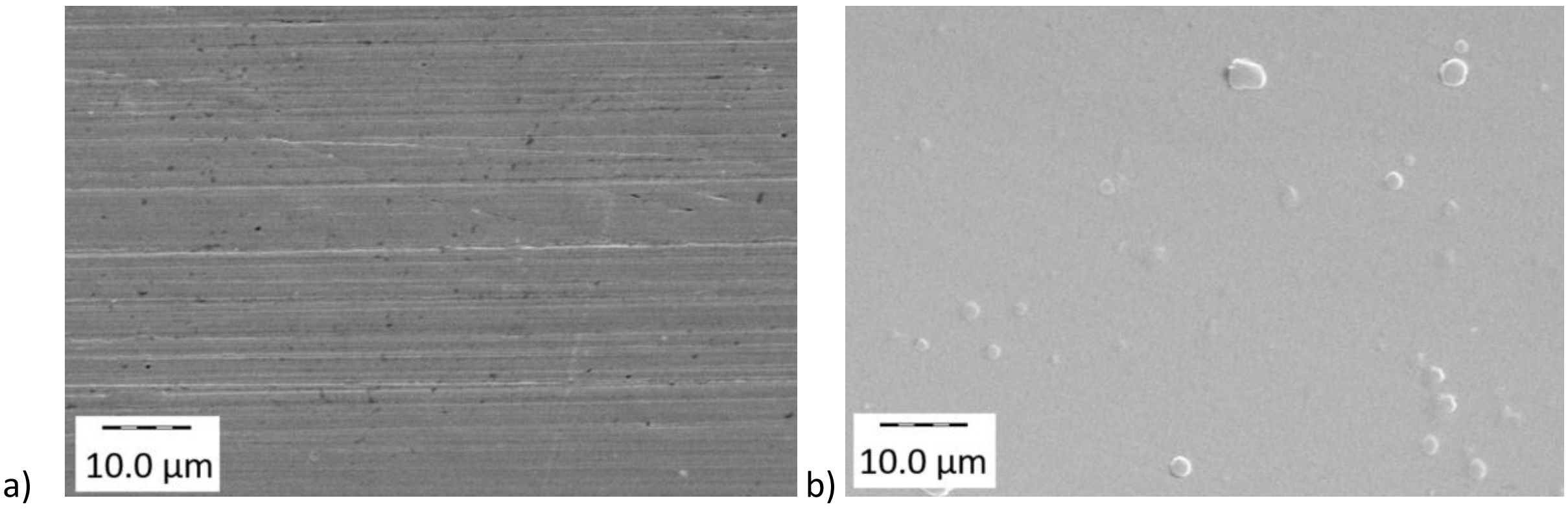

*Figure 16 SEM images of the surface of a) Ag, b) Ag/$W_{0.76}Ti_{0.24}B_{2.5}$ - after Tribotouch test.*

### 3.3 Corrosion resistance

Based on the open circuit potential test, the dependence of the corrosion potential on time was determined (Figure 17). Apart from the Ag/WB bilayer, the lowest values were recorded at the beginning of the test. The potential initially increases and then stabilises. Such curves indicate the transition of the material from an active to a passive state. The pure silver coating has a corrosion potential of 28 mV. After applying the WTiB layer, the potential decreased slightly; both materials are characterised by good corrosion resistance. In the case of the Ag/WB layer, the potential initially decreased and then stabilised. The lowest potential value of -560 mV was obtained, so it can be concluded that these materials have the lowest corrosion resistance.

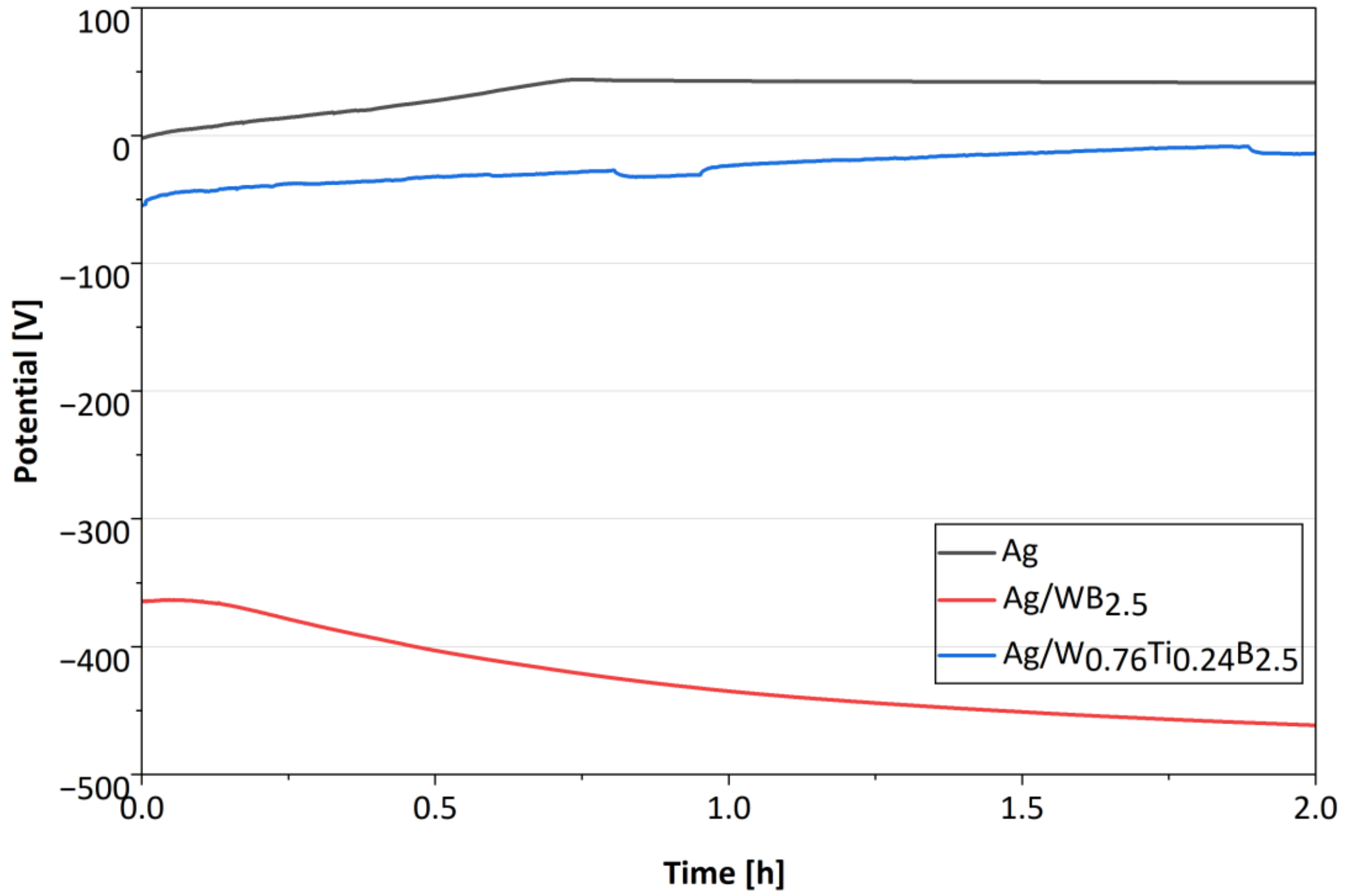

*Figure 17 Open circuit potential as a function of time.*

A further study to characterise the electrochemical behaviour of the layers was conducted using potentiodynamic polarisation. Figure 18 shows Tafel plots. Based on the analysis of the curves, characteristic electrochemical values were determined, i.e. corrosion current density ($i_{cor}$), corrosion potential ($E_{cor}$), polarisation resistance ($R_{pol}$) (Table 5). The Ag/WB layer has the highest corrosion current density ($i_{cor}$ = 7.33·10$^{-10}$ A/cm2), the lowest corrosion potential ($E_{cor}$ = -560 mV) and low polarisation resistance ($R_{pol}$ = 8.18·10$^{7}$ Ω*cm2). The graph does not show a clear passivation area; the surface is active and susceptible to dissolution in the electrolyte. Materials with lower potential are more chemically active, which makes them more susceptible to corrosion. Therefore, the Ag/WB layer has the lowest corrosion resistance among the tested samples.

In the initial stage of polarisation of the pure silver layer from the corrosion potential upwards, a narrow passivation area can be seen, which is associated with the formation of a protective film. This sample obtained the highest corrosion potential ($E_{cor}$ = 28 mV) and polarisation resistance ($R_{pol}$ = 2.33·10$^{8}$ Ohm*cm$^2$), as well as the lowest current density ($i_{cor}$ = 1.32·10$^{-10}$ A/cm$^2$), which indicates its highest corrosion resistance. The shift of the corrosion potential ($E_{cor}$) in the positive direction and the reduction in corrosion current density ($i_{cor}$) indicate a clear increase in corrosion resistance. On the other hand, the high polarisation resistance value, which contributes to a decrease in the anodic current, reduces the rate of corrosion [52]. After applying a titanium-doped tungsten boride coating, the corrosion potential shifts slightly towards lower values compared to pure silver. Therefore, both materials retain good corrosion properties.

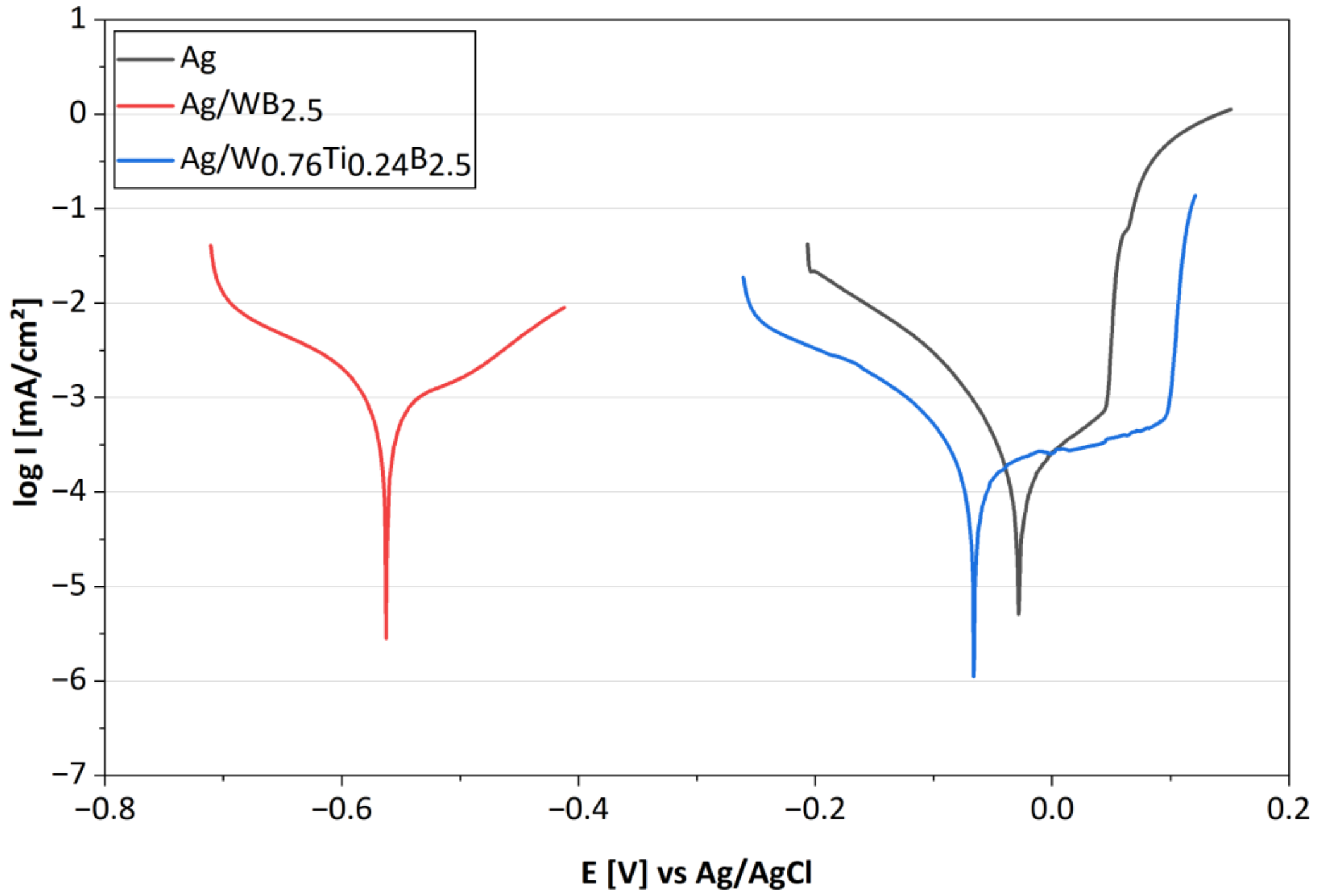

*Figure 18 Tafel curves.*

*Table 5 Electrochemical characteristic values obtained by anodic polarization.*

| Sample | $I_{cor}$ [A/cm²] | $E_{cor}$ [mV] | $R_{pol}$ [Ohm*cm²] |
|---|---|---|---|
| **Ag** | $1.32 \cdot 10^{-10}$ | -28 | $2.33 \cdot 10^{8}$ |
| **Ag/$WB_{2.5}$** | $7.33 \cdot 10^{-10}$ | -560 | $8.18 \cdot 10^{7}$ |
| **Ag/$W_{0.76}Ti_{0.24}B_{2.5}$** | $1.61 \cdot 10^{-10}$ | -63 | $3.75 \cdot 10^{8}$ |

## 4. Summary and conclusions:

The study developed a new bilayer consisting of silver and transition metal borides. An innovative method combining PLD laser deposition and HIPIMS magnetron sputtering was used. A series of tests were conducted to verify the functional properties and microstructure of the Ag, Ag/$WB_{2.5}$ and Ag/$W_{0.76}Ti_{0.24}B_{2.5}$ layers. Thanks to the appropriate approach, the disadvantage of the PLD method, i.e. the high content of droplets on the surface of the layers, was turned into an advantage and offers a significant improvement in wear resistance. The following conclusions can be drawn from the obtained test results:

- Ag/WTiB bilayer exhibit more than 10 times higher hardness and increased abrasion resistance compared to silver coatings. At the same time, the addition of titanium significantly increases the material's resistance to cracking. Coatings with titanium exhibit significantly higher adhesion to the substrate and reinforcement with silver particles. In addition, this composite has very good corrosion resistance, similar to pure silver. It has high corrosion potential and an order of magnitude higher polarisation resistance compared to the Ag/$WB_{2.5}$ bilayer.
- According to diffraction analysis, Ag/$WB_{2.5}$ coatings is composed of crystalline silver droplets and an amorphous/crystalline matrix. This material achieved very high hardness (22.06 ± 3.03 GPa) while exhibiting good anti-corrosion properties. Unfortunately, the coating is too brittle and poorly bonded to the substrate, resulting in brittle cracking and delamination under load.
- Tribotouch tests provided important information on the behaviour of the tested surfaces during everyday use. The Ag/WTiB bilayer has three times better anti-wear properties compared to pure silver coatings under operating conditions. Combined with the expected very good antibacterial properties, they are a promising option for coating everyday objects where high reliability and sterility are required. On the other hand, they can also be used as a protective coating for highly conductive microchips made of silver.

**Originality statement**

I write on behalf of myself and all co-authors to confirm that the results reported in the manuscript are original and neither the entire work, nor any of its parts have been previously published. The authors confirm that the article has not been submitted to peer review, nor has been accepted for publishing in another journal. The author(s) confirms that the research in their work is original, and that all the data given in the article are real and authentic. If necessary, the article can be recalled, and errors corrected.

**Declaration of competing interest**

The authors declare that they have no known competing financial interests or personal relationships that could have appeared to influence the work reported in this paper.

**Funding sources**

This research did not receive any specific grant from funding agencies in the public, commercial, or not-for-profit sectors.

**Acknowledgements**

This work was supported by the National Science Centre (NCN - Poland) Research Project: 2022/47/B/ST8/01296.

The authors would like to thank Prof. Marcin Pisarek (Institute of Physical Chemistry, Polish Academy of Sciences) for his assistance with additional measurements.

**CRediT authorship contribution statement**

Katarzyna Zielińska Conceptualization, Methodology, Investigation, Writing – original draft
Mateusz Włoczewski Investigation
Rafał Psiuk Writing – review and editing
Jacek Hoffman Investigation
Ewa Wojtiuk Investigation
Piotr Bazarnik Investigation
Tomasz Mościcki Conceptualization, Writing – review and editing, Supervision